\documentstyle[12pt,aaspp4]{article}
\newcommand{\be}{\begin{equation}}
\newcommand{\ee}{\end{equation}}
\begin{document}
{\hfill Preprint}
\title{Reconnection in a Weakly Stochastic Field}

\author{ A. Lazarian\altaffilmark{1,3} and Ethan T. Vishniac\altaffilmark{2}}
\altaffilmark{1}{Princeton University Observatory, Princeton, NJ 08544;
lazarian@astro.princeton.edu}

\altaffilmark{3}{CITA, University of Toronto, Canada; 
lazarian@cita.utoronto.ca}

\altaffilmark{2}{Department of Physics and Astronomy, Johns Hopkins University, 
Baltimore MD 21218; ethan@pha.jhu.edu}

\begin{abstract}
We examine the effect of weak, small scale magnetic field structure on
the rate of reconnection in a strongly magnetized plasma.  
This affects the rate of reconnection by reducing the 
transverse scale for reconnection flows, and by allowing
many independent flux reconnection events to  
occur simultaneously.  Allowing only for the first effect
and using Goldreich and Sridhar's model of strong turbulence in a 
magnetized plasma with negligible intermittency, we find a
lower limit for
the reconnection speed $\sim V_A {\cal R}_L^{-3/16}{\cal M}^{3/4}$, 
where $V_A$ is the Alfv\'en speed, ${\cal R}_L$ is the 
Lundquist number, and ${\cal M}$ is the large scale magnetic Mach number of
the turbulence.  We derive an upper limit of $\sim V_A{\cal M}^2$
by invoking both effects.  We argue that 
generic reconnection in turbulent plasmas will normally occur at close to
this upper limit. The fraction of magnetic energy that
goes directly into electron heating scales as 
${\cal R}_L^{-2/5}{\cal M}^{8/5}$,
and the thickness of the current sheet scales as 
${\cal R}_L^{-3/5}{\cal M}^{-2/5}$.
A significant fraction of the magnetic energy goes into high frequency Alfv\'en waves.
The angle between adjacent field lines on the same side of the
reconnection layer is $\sim {\cal R}_L^{-1/5}{\cal M}^{6/5}$ on the scale of the
current sheet thickness.
We claim that the qualitative sense of these conclusions, that
reconnection is fast even though current sheets are narrow, is
almost independent of the local physics of reconnection and
the nature of the turbulent cascade. As the consequence of this
the Galactic and Solar dynamos are generically fast, i.e. do not
depend on the plasma resistivity. 
\end{abstract}


\section{Introduction}

Magnetic reconnection is a long standing problem in
theoretical magnetohydrodynamics (MHD). This problem is closely
related to the hotly debated problem of the magnetic dynamo 
(see Parker 1979, Moffatt 1978, 
Krause \& Radler 1980). Indeed, it is impossible to understand the 
origin and evolution of large scale magnetic fields without
a knowledge of the mobility and reconnection of magnetic fields.
In a typical astrophysical plasma, resistivity is very small
and flux freezing should be an excellent approximation to the 
motion of magnetic field, while dynamo action invokes a
constantly changing magnetic field topology
\footnote{Merely winding up a magnetic field can increase
the magnetic field energy, but cannot increase the magnetic
field flux. We understand the dynamo in the latter sense.
The Zel'dovich ``fast'' dynamo (Vainshtein \& Zel'dovich 1972) also
invokes reconnection for continuous dynamo action
(Vainshtein 1970).} This means an astrophysical dynamo requires 
efficient reconnection despite very slow Ohmic diffusion rates.

The literature on magnetic reconnection is rich and vast (see, for
example, Biskamp 1993 and references therein). 
In the simplest model of magnetic reconnection (\cite{P57}, \cite{S58},
see \S 3.1) the reconnection speed lies below the Alfv\'en speed
by a factor
$\sim {\cal R}_L^{-1/2}=(\eta/V_A L_x)^{1/2}$, where 
$\eta$ is the resistivity, $V_A$ is the Alfv\'en speed, and $L_x$
is the length of the current sheet, assumed to be determined
by the large scale geometry of the problem.  The large ratio of
$L_x$ to the width of the current sheet, $\Delta$, equal to the ratio of
the resistivity to the reconnection speed, drives the reconnection
speed towards zero for highly conducting fluids.  In particular,
this is many
orders of magnitude too small for astrophysical reconnection.
Consequently,
there has been intense interest in models which result in higher
reconnection speeds.  

In general, we can divide schemes for fast reconnection into those
which alter the microscopic resistivity, broadening the current sheet,
and those which change the global geometry, thereby reducing $L_x$. 
An example of the latter is the suggestion by 
Petschek (1964) that reconnecting magnetic
fields would tend to form structures whose typical size in
all directions is determined by the resistivity (`X-point' reconnection).  
This results in
a reconnection speed of order $V_A/\ln {\cal R}_L$.  On the other
hand, attempts to produce such structures in simulations
of reconnection have been disappointing (\cite{B84}, \cite{B86}).
In such simulations the X-point region typically collapses towards the
Sweet-Parker geometry as the Lundquist number becomes
large (\cite{B96}, Wang, Ma \& Bhattacharjee 1996, Wang \& Bhattacharjee
1996).  One way to understand this collapse is to consider perturbations
of the original X-point geometry.  In order to maintain this geometry
reconnection has to be fast, which requires shocks in the original
(Petschek) version of this model.  These shocks are, in turn, supported
by the flows driven by fast reconnection, and fade if $L_x$ increases.
Naturally, the dynamical range for which the existence of such shocks
is possible depends on the Lundquist number and shrinks when
fluid conductivity increases.  The apparent conclusion is that 
at least in the collisional regime
reconnection occurs through narrow current sheets.
\footnote{A very different approach explored by Vishniac (1995a,b) 
appealed to the concentration of magnetic field in narrow flux
tubes within a high $\beta$ plasma, thereby reducing $L_x$ to
the flux tube radius. Given the
small filling factor of the magnetic field and the high Alfven velocity
inside the flux tubes it is possible to obtain fast reconnection even using
the Sweet-Parker geometry. However, flux tube
formation requires initially high reconnection rates.}

In the collisionless regime the width of the current sheets
may be determined by the ion cyclotron (or Larmor) radius
(\cite{P79}) or by the ion skin depth (\cite{MB96},
\cite{BSD97}, \cite{SDDB98}).
This often leads to a current sheet thickness which is much larger
than expected (`anomalous resistivity').  Furthermore,  anomalous
resistivity can be combined with X-point reconnection with the
result that the X-point is somewhat less unstable to collapse.
Nevertheless,  while this effect can dramatically increase reconnection speeds
in simulations and in laboratory experiments it does not
change the qualitative conclusion that narrow current sheets
give reconnection speeds in astrophysical plasmas that are
unacceptably low.
\footnote{Reconnection in the
presence of ambipolar diffusion discussed in Vishniac \& Lazarian (1998)
vaguely fits into the category of ``enhanced $\Delta$''. Indeed,
the neutrals do not feel magnetic field and leave the reconnection zone
forming an outflow that is limited by ambipolar diffusion rather than 
Ohmic diffusivity.}

In order to explain why astrophysical magnetic fields do not reverse on
very small scales, researchers have usually appealed to 
an {\it ad hoc} diffusivity 
which is many orders of magnitude greater than Ohmic
diffusivity.  This diffusivity is assumed to be roughly
equal to the local turbulent diffusion coefficient.
While superficially reasonable, this choice implies that
a dynamically significant magnetic field diffuses through
a highly conducting plasma in much the same way as a 
tracer of no dynamical importance
\footnote{In the case of galaxies the ratio of the two
diffusivities is $\sim 10^{20}$.}. This is referred to as
turbulent diffusivity and denoted $\eta_t$, as opposed
to the Ohmic diffusivity $\eta$. Its name suggests
that turbulent motions subject the field
to kinematic swirling and mixing. As the field becomes
intermittent and intermixed it can be assumed to undergo dissipation at
arbitrarily high speeds.

Parker (1992) showed convincingly that the concept of turbulent
diffusion is ill-founded. He pointed out that turbulent motions are strongly
constrained by magnetic tension and large scale magnetic fields
prevent hydrodynamic motions from mixing magnetic field regions of opposing 
polarity unless they are precisely anti-parallel.  Both numerical and 
analytic studies 
(see \cite{CV91}, \cite{KA92}, \cite{GD94}) confirm that the 
traditional (\cite{RSS88})
theory of kinematic dynamos is seriously and fundamentally flawed. 
On the other hand, observations of the solar corona and chromosphere
seem to show that reconnection often takes place at speeds of
$\sim 0.1V_A$ (cf. Dere 1996, Innes et al. 1997 and references contained
therein).  Evidently at least some astrophysical plasmas can undergo 
reconnection on short time scales, regardless of the 
apparent
theoretical
difficulties involved.

An associated problem for dynamo theories is the growth of small scale random
magnetic fields. Assuming magnetic fields are allowed to intermix freely,
as the turbulent diffusion model requires,
Low (1972) and Krause \& Radler (1980) estimated that the
ratio of the mean field to the random field is expected to be 
of the order $\sqrt{\eta_t/\eta}\gg 1$. At the same time, observations
indicate that these field components are 
not very different
in 
galaxies at the present epoch (Beck et al.\ 1996). 
{}Furthermore, Kulsrud and Anderson (1992) have suggested that the growth
of a random field should suppress dynamo action. 

This unsatisfactory state of affairs has stimulated recent attempts to
find new approaches to the problem of the astrophysical 
dynamo\footnote{The astrophysical dynamo is just an example of a process
whose dynamics is determined by reconnection rates. Solar flaring
(see Dere 1996) and star formation in molecular clouds (for example
see Lubow \& Pringle 1996) are other examples.}
(e.g. Parker
1992, Parker 1993, Vishniac 1995a,b, Kulsrud et al 1997) that
have deficiencies which we will discuss in more detail below.

Below we explore a different approach to the problem of rapid reconnection.
This paper deals with the issue of magnetic reconnection\footnote{The
mode of reconnection discussed here is sometimes is called {\it free} 
reconnection
as opposed to {\it forced} reconnection. Wang \& Bhattacharjee (1992)
define {\it free} reconnection as a process caused by a nonideal instability
driven by the free energy stored in an equilibrium. If
the equilibrium is stable, reconnection can be forced if a 
perturbation is applied externally.} 
in the presence
of a weak random field component.   
Magnetohydrodynamic turbulence guarantees the presence of such a field, 
although its amplitude and structure clearly depends on the model we adopt 
for MHD turbulence, as well as the specific environment of the field. 
It is also obvious from the above discussion that as long as the kinematic 
approximation holds, that is as long as magnetic forces are negligible,
\footnote{This approximation should be valid, at least, at the initial stages
of magnetic field generation.} the
random component of the magnetic field will quickly come to 
dominate the large scale one.  We will assume throughout that we can
describe resistive effects with an Ohmic resistivity $\eta$.  However,
our qualitative conclusions are not sensitive to this choice.
We will show that if we allow for the more complicated geometry
expected from a turbulent cascade, the speed of reconnection 
is fast, independent of the plasma resistivity, and a function
only of the overall level of turbulence.  Although we will adopt
a specific model of MHD turbulence in this work, we will also show
that our conclusions are not sensitive to this choice.

There are two mechanisms which lead to this result.  First, if
we consider two adjoining regions with very different large scale
fields, the small scale structure of both fields implies that
the typical scale for a reconnection event is much smaller than
the overall size of the system, and goes to zero in the limit of
an ideal fluid.  Second, since field lines wander in and out of the 
reconnection zone, the overall rate of flux reconnection can be
much greater than the rate for a single event.  We note that
it is usually believed that for reconnection to be rapid
in the limit of $\eta \rightarrow 0$ a current singularity should
develop (Park et al 1984). Our model does not require such singularities.
Indeed, we show that while the amount of Ohmic dissipation tends to 0 as
$\eta \rightarrow 0$, the magnetic field has a weak stochastic 
substructure on the smallest
scales and the rate of the flux reconnection does not decrease.

In \S 2 of this paper we will introduce the scheme of 
reconnection in the presence of
stochastic field, while in \S 3
of this paper we will describe the results of a simple model 
for the small scale structure of a large scale magnetic field embedded in a 
turbulent medium.  A fuller discussion of this model is given in
Appendix A. 
In \S 4 we will estimate the reconnection
speed in the presence of small amounts of stochasticity, and apply
these estimates to the model described in \S 3.  In \S 5 we 
describe some of the implications of this work.
We give our conclusions in \S 6.

\section{The Problem}

Sweet-Parker reconnection (Sweet 1958, Parker 1957)
provides a minimum value for reconnection rates. In this model
oppositely directed magnetic fields are brought into contact
over a region of size $L_x$. Magnetic fields reconnect along a
very thin Ohmic diffusion layer $L_y\approx \eta/V_{rec}$ and fluid
is ejected from this layer at a velocity of the order $V_A$ in 
a direction parallel to the local field lines.  The layer
in which Ohmic diffusion takes place is usually referred
to as the current sheet.  Here we will refer to the volume
where the mean magnetic field strength drops significantly
as the reconnection zone, in order to allow for the presence
of collective effects which may broaden the reconnection zone
well beyond the current sheet. 
The reconnection velocity in the Sweet-Parker
picture is determined by the constraint imposed by the conservation of mass
condition $V_{rec}L_x\approx V_A L_y$.  Although this model
is two dimensional, it can be generalized to three dimensions
by allowing the two magnetic field regions to share a common
field component, which has the effect of rotating them so
that they are no longer exactly anti-parallel.  This has
no effect on the Sweet-Parker reconnection process (see Fig.~1).
However, it does change the nature of the constraint somewhat.
In addition to ejecting matter from the reconnection zone, we
must also allow for the ejection of the magnetic flux due to the
common field component.  This is, in effect, the same constraint
in this case.

We consider the case in which there exists a large scale,
well-ordered magnetic field, of the kind that is normally used as
a starting point for discussions of reconnection.  This field may,
or may not, be ordered on the largest conceivable scales.  However,
we will consider scales smaller than the typical radius of curvature
of the magnetic field lines, or alternatively, scales below the peak
in the power spectrum of the magnetic field, so that the direction
of the unperturbed magnetic field is a reasonably well defined concept.
In addition, we expect that the field has some small scale `wandering' of
the field lines.  On any given scale the typical angle by which field
lines differ from their neighbors is $\phi\ll1$, and this angle persists
for a distance along the field lines $\lambda_{\|}$ with
a correlation distance $\lambda_{\perp}$ across field lines.  

The modification of the mass conservation constraint in the presence of 
stochastic magnetic field component 
is self-evident. Instead of being squeezed from a layer whose
width is determined by Ohmic diffusion, the plasma may diffuse 
through a much broader layer, $L_y\sim \langle y^2\rangle^{1/2}$ (see Fig.~2), 
determined by the diffusion of magnetic field lines. The value of
$\langle y^2\rangle^{1/2}$ can be determined once a particular model
of turbulence is adopted (see \S 3), but it is obvious from the very beginning
that this value is determined by field wandering rather than Ohmic
diffusion as in the Sweet-Parker case. 

In the presence of a stochastic field component, magnetic reconnection 
dissipates field lines not over their  entire length $\sim L_x$ but only over
a scale $\lambda_{\|}\ll L_x$ (see Fig.~2b), which
is the scale over which magnetic field line deviates from its original
direction by the thickness of the Ohmic diffusion layer $\lambda_{\perp}^{-1}
\approx \eta/V_{rec, local}$. If the angle $\phi$ of field deviation
does not depend on the scale, the local
reconnection velocity would be $\sim V_A \phi$ and would not depend
on resistivity. We claim in \S 3 that $\phi$ does depend on scale. 
Therefore the {\it local} 
reconnection rate $V_{rec, local}$ is given by the usual Sweet-Parker formulae
but with $\lambda_{\|}$ instead of $L_x$, i.e. $V_{rec, local}\approx V_A 
(V_A\lambda_{\|}/\eta)^{-1/2}$.
It is obvious from Fig.~2a that $\sim L_x/\lambda_{\|}$ magnetic field 
lines will undergo reconnection simultaneously (compared to a one by one 
line reconnection process for
the Sweet-Parker scheme). Therefore the overall reconnection rate
may be as large as
$V_{rec, global}\approx V_A (L_x/\lambda_{\|})(V_A\lambda_{\|}/\eta)^{-1/2}$,
which means that the reconnection efficiency critically depends on
the value of $\lambda_{\|}$.  More realistically we will find that
there are other global constraints which end up determining the
actual global reconnection speed.

The relevant values of $\lambda_{\|}$ and $\langle y^2\rangle^{1/2}$ 
depend critically
on the magnetic field statistics. Therefore in the next section we will
briefly explore the expected properties of magnetic turbulence.

\section{Magnetic Field Structure in a Turbulent Medium}

The study of MHD turbulence is hampered by the absence of 
easily observable examples.  We will base our discussion
here on a model first proposed by Goldreich and Sridhar 
(1995, hereinafter GS95),
which is at least consistent with available observations
of the interstellar medium and the solar wind.  A discussion
of this model, and the justification for using it as a
first approximation is given in Appendix A.
In any case, the qualitative
nature of our results, that a weak stochastic component to the
field structure can have a dramatic effect on reconnection rates,
is not sensitive to the details of the model we adopt.
We discuss reconnection for various power spectra including
the Kraichnan spectrum (Iroshnikov 1963, Kraichnan 1965) in
Appendix D.

{}For strong turbulence we have 
$\lambda_{\perp}\approx \lambda_{\|}\phi(\lambda_{\|})$,
although we can also define a weakly turbulent regime where 
the correlation length is much greater than the distance by
which individual field lines deviate from a straight line. 
In either case, 
we can define a corresponding perpendicular
wavenumber, $k_{\perp}\sim\lambda_{\perp}^{-1}(\lambda_{\|})$.

In order to describe this wandering, we need to know something about
quasi-static distortions of the field lines, or `zero-frequency' modes.
Normally, this means only those distortions of the
magnetic field whose evolution is driven
primarily by nonlinear effects.  Here, however, we mean simply
all modes that evolve on time scales
long enough that they can be considered as a background to the
evolution of flows driven by reconnection. 
Motions that can be described as Alfv\'en waves may be relevant 
in this sense, as long as they involve very long wavelengths
parallel to the magnetic field.  However, most of our discussion
will be concerned with modes with nonlinear time scales
comparable to their wave periods, so that they are Alfv\'en
waves only in a very general sense.
{}For simplicity, we will
consider incompressible turbulence and assume that the magnetic field
filling factor is not too small.
The former should be a reasonable 
approximation for small scale, and therefore weak, fluid motions.  
The latter is probably a bad approximation inside the Sun, although
reasonable whenever the magnetic pressure is comparable to, or greater than,
the gas pressure (or when intermittency has not had time to develop).  

We can describe a model of MHD turbulence in terms of the relationship
between $k_{\perp}$ and $k_{\|}$ at every scale in the turbulent 
cascade, as well as the scaling of $v_k$ as function of wavenumber. 
By restricting ourselves to a model in which field stochasticity
is driven locally by a turbulent cascade we are neglecting the 
possibility that the 
field lines are anchored in some denser medium, like field lines extruded
from the photosphere of the Sun, but even in this case the field lines
are mixed at their footpoints by turbulent motions in the photosphere
(see, for example, Parker 1988, and references contained therein).
More generally there will be cases in which the 
level of stochasticity will be determined by nonlocal processes.
Here we will restrict ourselves to cases where the field stochasticity
can be understood in terms of local turbulence.

If energy is injected on some scale $l$, with $v_l\le V_A$, then
GS95 predict that 
\be
k_{\|} \approx l^{-1} (k_{\perp}l)^{2/3}
\left({v_l\over V_A}\right)^{4/3},
\label{k_p2m}
\ee
\be
\tau_{nl}^{-1}\approx {v_l\over l}(k_{\perp}l)^{2/3}\left({v_l\over V_A}\right)^{1/3},
\label{tau2m}
\ee
while the rms fluid velocity is given by
\be
v_k\approx v_l (k_{\perp}l)^{-1/3}\left({v_l\over V_A}\right)^{1/3}.
\label{velm}
\ee
These equations are derived in Appendix A.
Equation (\ref{k_p2m}) is particularly interesting. It 
gives the geometry of small scale magnetic
field structure in a turbulent cascade.  This turns out to be
critically important in magnetic reconnection.

A bundle of field lines confined within a region of width $y$
at some particular point will spread out perpendicular to the mean
magnetic field direction as one moves in either direction parallel
to the local magnetic field lines.  The rate of field line diffusion
is given approximately by
\be
{d\langle y^2\rangle\over dx}\sim {\langle y^2\rangle\over \lambda_{\|}},
\ee
where $\lambda_{\|}^{-1}\approx k_{\|}$, $k_{\|}$ is the parallel 
wavevector chosen so that the corresponding vertical wavelength, $k_{\perp}(k_{\perp})$,
is $\sim \langle y^2\rangle^{1/2}$, and $x$ is the
distance along an axis parallel to the mean magnetic field.
Therefore, using equation
(\ref{k_p2m}) one gets
\be
{d\langle y^2\rangle\over dx}\sim l\left({\langle y^2\rangle\over l^2}\right)^{2/3}
\left({v_l\over V_A}\right)^{4/3} 
\label{eq:diffuse}
\ee
where we have 
substituted $\langle y^2\rangle ^{-1/2}$ for $k_{\perp}$.  This expression for the
diffusion coefficient will only apply when $y$ is small enough for us
to use the strong turbulence scaling relations, or in other words when
$\langle y^2\rangle < l^2(v_l/V_A)^4$.  Larger bundles will diffuse at a maximum rate
of $l(v_l/V_A)^4$.  For $\langle y^2\rangle$ small equation (\ref{eq:diffuse}) implies
that a given field line will wander perpendicular to the mean field
line direction by an average amount
\be
\langle y^2\rangle^{1/2}= {(3x)^{3/2}\over l^{1/2}} \left({v_l\over V_A}\right)^{2}
\label{eq:diffuse2}
\ee
in a distance $x$.  The fact that the rms perpendicular displacement grows
faster than $x$ is significant.  It implies that if
we consider a reconnection zone, a given magnetic flux element that
wanders out of the zone has only a small probability of wandering
back into it.

\section{Turbulent Reconnection}

\subsection{Constraints on reconnection rate}

Outflow of matter from the reconnection layer constrains the achievable
reconnection rates. In the presence of turbulence
the thickness of the outflow layer increases with $L_x$ according
to equation (\ref{eq:diffuse2}): 
\be
\langle y^2\rangle^{1/2}\sim L_x\left({L_x\over l}\right)^{1/2}
\left({v_l\over V_A}\right)^{2},
\label{eq:diff2}
\ee
when $l>L_x$ and 
\be
\langle y^2\rangle^{1/2}\sim \left(L_x l\right)^{1/2}
\left({v_l\over V_A}\right)^{2},
\label{eq:diff3}
\ee
when $L_x>l$.
Therefore the upper limit on $V_{rec}$ imposed by large scale
field line diffusion is 
\be
V_{rec}<V_A\min\left[\left({L_x\over l}\right)^{1/2},
\left({l\over L_x}\right)^{1/2}\right]
\left({v_l\over V_A}\right)^{2}.
\label{eq:lim2a}
\ee
This limit on the reconnection speed is
fast, both in the sense that it does not depend on the
resistivity, and in the sense that it represents a large
fraction of the Alfv\'en speed.  Whether or not this
limit is attainable depends to a large extent on the motion
of the shared component of the magnetic flux.  
Nevertheless, we shall see below that
equation (\ref{eq:lim2a}) is not
merely an upper limit on the global reconnection speed, but
under most circumstances, constitutes a reasonable
estimate for its actual value.

There are two important issues that we have neglected
in arriving at this estimate.  First, we have assumed an
outflow velocity of $V_A$ spread over the entire width
of the outflow region.  However, in general the field lines
that pass through the reconnection zone at any one moment
will not fill the entire width of the outflow region.  
Instead, they will fill some fraction of it, with a 
proportionate drop in the effective outflow velocity.   
This reduction will be almost exactly offset by the fact that 
every flux element stays in the reconnection zone
for roughly $L_x/V_A$ and that defines the meaning of 
``simultaneous'' in this case. When reconnection starts,
the outflow fills only a small fraction of the reconnection 
layer and this corresponds to the minimal estimate of the
reconnection rate below. However, as reconnection proceeds,
more independent flux elements become involved and this
enhances the rate. 

In addition, equation (\ref{eq:lim2a}) is based on the assumption that
the field line topology far from the reconnection layer
is not affected by the process of reconnection.  In 
reality we expect that, once reconnection begins,
the changing field line stresses will be communicated throughout
the volume at a speed $V_A$.  Also,
the reconnection layer will act as a source of small scale 
nonlinear Alfv\'en waves.  The outcome of these effects is
somewhat uncertain, but it will probably have the effect
of accelerating reconnection.

We can obtain a lower limit on the reconnection speed by
narrowing our focus and remembering 
that plasma can escape from the reconnection layer proper only by 
traveling a distance $\lambda_{\|}(k_{\bot})$ along the local
field lines, where
$k_{\bot}^{-1}\sim \Delta$ is the thickness of the reconnection layer.
A lower limit on the reconnection speed follows from the mass
conservation condition:
\be
V_{rec,local}\lambda_{\|}\approx V_A \Delta,
\label{V_r1}
\ee
and the requirement that at a fundamental level reconnection
is driven by Ohmic diffusion:
\be
V_{rec,local}\approx \eta/\Delta.
\label{V_r2}
\ee
We have denoted this estimate for the reconnection speed with a 
special subscript not only because its value is 
determined by local physics, but also because we shall see that
it is likely that the global speed of reconnection through
a field line bundle is different, and much faster, than 
this local value.
Combining equations (\ref{k_p2m}), (\ref{V_r1}) and (\ref{V_r2}) and recalling
that $k_{\bot}^{-1}\sim \Delta$ and $k_{\|}^{-1}\sim \lambda_{\|}$
we get reconnection at a speed
\be
V_{rec,local}\sim {V_A\over k_{\perp}\lambda_{\|}}\sim V_A(k_{\perp}l)^{-1/3}
\left({v_l\over V_A}\right)^{4/3}\sim 
V_A\left({V_{rec,local} l\over\eta}\right)^{-1/3} 
\left({v_l\over V_A}\right)^{4/3},
\ee
which implies that
\be
V_{rec,local}\sim v_l\left({\eta\over V_Al}\right)^{1/4}.
\label{eq:vrec1}
\ee
In arriving at this estimate we have treated field perturbations
with frequencies less than $\sim V_A/\lambda_{\|}$ as
essentially static.  However, since we see from equations
(\ref{V_r1}) and (\ref{V_r2}) that the reconnection rate
$\eta/\Delta^2$ is $\sim V_A/\lambda_{\|}$, this is
a reasonable approximation.  Finally, we note that this process is
controlled by field fluctuations with a perpendicular wavenumber
\be
k_{\perp}\approx {V_{rec,local}\over \eta}\approx l^{-1}{v_l\over V_A}
\left({V_A l\over \eta}\right)^{3/4}.
\label{eq:kskin}
\ee
We see from equation (\ref{eq:eta}) that this is the largest wavenumber
we can expect in the turbulent cascade, or at least the largest we can
expect before the reconnection process generates its own local turbulence.


If, however, the medium is partially ionized the maximum wavenumber 
given in equation (\ref{eq:eta}) should be reduced by a factor 
$\sim (\eta/\eta_{amb})^{3/4}$. 
 This will reduce the reconnection
speed by a factor of $\sim (\eta/\eta_{amb})^{1/4}$, but this 
reduction will be somewhat offset by the fact that the ions will be 
compressed until their pressure alone is comparable to the magnetic field pressure, 
and it is this compressed layer of ions which is ejected from the reconnection
region along the field lines.  The resulting velocity should
be compared with the ambipolar reconnection velocity from (Vishniac
\& Lazarian 1998):
\be
V_{ambipolar}=\left(\frac{2\eta}{\tau_{recomb}}\right)^{1/2}
\frac{1}{\beta X}(1+2\beta x)^{1/2}
\label{eq:ambip0}
\ee
whether $X$ is the ionization ratio, $\beta$ is the ratio of the 
gas pressure to the magnetic pressure. The 
recombination time of the ions $\tau_{recomb}$ as well as Ohmic diffusivity
$\eta$ are taken far gas from the reconnection layer. The larger of
the two velocities will determine the reconnection speed. Results
in Table~1 in Vishniac \& Lazarian (1998) show that the reconnection
speed can be increased via ambipolar diffusion by a factor of $10^3$, 
which is not much, if we account for the fact that the original
Sweet-Parker rate is very slow. 
A detailed account of reconnection in the presence of MHD turbulence
and ambipolar diffusion is given in Lazarian \& Vishniac (1998).  

So far we have ignored the fact that the thin current sheets envisioned
in Sweet-Parker reconnection are unstable.  We briefly address this
issue in Appendix C and show that allowing for the presence of tearing
modes raises the minimum value of $V_{rec,local}$ to  
$\sim V_A \left({v_l\over V_A}\right)^{3/4} \left({\eta\over V_Al}\right)^{3/16}$.
Even in the context of large scale turbulence in the interstellar
medium, this is only an increase of about one order of magnitude,
and still leaves us with $V_{rec,local}\ll V_A$.

\subsection{Estimating the global value of $V_{rec}$}

The physics of the reconnection
layer will be significantly more complicated than the simple
picture above indicates. First of all, a number of known 
processes can enhance reconnection. In the appendices we discuss
Bohm diffusion, anomalous resistivity and tearing modes.
All of these have been discussed previously as ways to 
accelerate reconnection, but none have emerged as universal
mechanisms for producing fast reconnection.
In a previous paper (Vishniac and Lazarian 1998) we have shown 
that while ambipolar diffusion can produce dramatically faster
reconnection rates, the overall pace of reconnection is still
orders of magnitude below $V_A$.
Anomalous resistivity (see Appendix B) is of marginal importance for
interstellar reconnection, although it can be important for laboratory
plasma. Finally, Bohm
diffusion (Appendix B) 
can provide a strong enhancement of reconnection speeds, but
unfortunately it is unclear whether or not the empirical
concept of Bohm diffusion is applicable to
astrophysical plasmas.  

All of this is probably less important than the global ordering of
reconnection events.  The minimal estimate of $V_{rec}$ given 
in the previous subsection is based on the assumption that reconnection
proceeds sequentially, that is, the reconnection speed is 
simply the speed with which reconnection propagates through
a single flux element.  This is not obviously correct, since the
reconnection zone contains many independent reconnection events
at any one time.  We need to define a global reconnection speed,  
$V_{rec,global}$, which describes the rate at which flux is reconnected
throughout the reconnection zone.  In order to arrive at a
reasonable estimate of this speed, we have to determine which
aspect of reconnection sets a limit on its efficiency.  Based
on our previous discussion, we can see that there
are four possibilities: the mass flow from the reconnection zone 
itself, the speed with which reconnected flux elements move across the
reconnection zone and off the edge, the ejection of the flux associated
with the shared magnetic field component, and 
the mass flow from the contact volume
(roughly everything within a distance $L_x$ of the reconnection
zone).  In the case of Sweet-Parker reconnection the first process
provides the critical constraint (and the third and fourth are not separate
constraints).

The first limit is determined by $V_{rec,local}$, but is not
equal to it.  Since the current sheet contains as many as
$k_{\perp}k_{\|} L_x^2$ independent reconnection surfaces,
each one reconnecting flux at a rate $\lambda_{\perp}^{-1}V_{rec,local}$,
the global reconnection speed could be enhanced by a factor\footnote{
The ambipolar reconnection rate given by Eq.~(\ref{eq:ambip0}) is
a local reconnection speed. Therefore, in the presence of turbulence
the global ambipolar reconnection speed is larger by a similar
factor (Lazarian \& Vishniac 1999).} 
of $k_{\|}L_x$.  In other words, using equations (\ref{k_p2m}), 
(\ref{eq:vrec1}), and (\ref{eq:kskin}) we get
\be
V_{rec,global}=k_{\|}L_x V_{rec,local}=V_A\left({V_Al\over\eta}\right)^{1/4}
{L_x\over l} \left({v_l\over V_A}\right)^3.
\label{gglob}
\ee
Since this will almost always be less restrictive than equation 
(\ref{eq:lim2a}),
we conclude that the flow of matter from the reconnection zone itself
does not limit the speed of reconnection. 

The second limit is set by the speed with which reconnected magnetic
field elements can pass through one another on their way to the edge
of the reconnection zone.  After initial reconnection magnetic field
elements will move until they encounter magnetic flux moving in
the opposite direction.  If this process limits the speed of
reconnection, then we can assume that each flux element has a typical
size comparable to the width of the reconnection zone, and moves a 
similar distance before being entangled by another flux element.
(If the reconnected flux elements were distributed more sparsely, then
from the preceding paragraph we can see that the gaps would fill in
with freshly reconnected flux.)  We can restrict ourselves to the
case where the flux elements cross the reconnection layer vertically,
rather than at a shallow angle, since in the latter case 
reconnection between opposing flux surfaces would, if anything, be
enhanced and the volume filled by a single reconnected element
would evolve into one filled by many flux elements with a nearly
vertical orientation.  Outside the reconnection layer the magnetic 
field moves parallel to the reconnection layer only a
small amount, and reverses direction after each interchange.  The
layers of magnetic field around the reconnection zone are removed
starting from its edges and working inwards.  

The steady pace of
reconnection ensures that the reconnection zone is in a constant
state of turbulent motion.  In the absence of 
collisions between flux elements we could assume an average
transverse velocity of $V_A$, but in their presence we need to
estimate the reconnection speed for these flux elements.  In
an important sense, this is just a repetition of the problem
we face in finding $V_{rec,global}$.  The only difference is that
turbulent motions within the reconnection zone guarantee a local
source of turbulence.  The level of turbulence in a reconnection
layer of width $\Delta$ is given by equating the energy injection rate, 
$V_A^2(V_{eject}/\Delta)$, with the turbulent dissipation 
rate, $\sim v_{local}^4/V_A\Delta$.
We conclude that
\be
v_{local}\sim V_A^{3/4}V_{eject}^{1/4}
\ee
We note that the limiting value
of $k_{\perp}$ imposed by dissipation outside the reconnection zone
will be replaced locally by some much larger value due to this separate
source of turbulence.  Consequently, if the motion
of flux elements in the reconnection zone limit the global
pace of reconnection, and if we adopt the ansatz\footnote{We do not consider
the possibility of a 
logarithmic dependence on the Lundquist number, as in this 
case reconnection is already fast.} that
\be
V_{rec,global}=V_A \left({\eta\over lV_A}\right)^n\left({v_l\over V_A}\right)^m,
\label{eq:vscale}
\ee
where $m$ and $n$ are to be determined, then the average ejection speed
of magnetic flux from the sides of the reconnection zone is just
\be
V_{eject}=V_A \left({\eta\over \Delta V_A}\right)^n
\left({V_{eject}\over V_A}\right)^{m/4}.
\label{eq:eject}
\ee
On the other hand, if reconnected magnetic flux elements fill a large
fraction of the reconnection zone, as they must if they determine the
overall pace of reconnection, then
\be
V_{rec,global}\approx V_{eject}.
\label{outflow}
\ee
We see from equations (\ref{eq:eject}) and (\ref{outflow}) that 
\be
V_{rec,global}\approx V_A\left({\eta\over\Delta V_A}\right)^{n\over1-(m/4)}.
\label{smallscale}
\ee 
We can see from this that $m<4$, but there is no obvious constraint on
$n$.  However, since the intersection of separate flux elements constitutes
a repetition, on a smaller scale, of the global process of reconnection, we
expect to find a still smaller scale $\Delta'$ which characterizes the
size of the flux elements generated during the reconnection of flux elements
of size $\Delta$.  The same arguments then allow us to conclude that
\be
V_{rec,global}\approx V_A\left({\eta\over\Delta' V_A}\right)^{n\over1-(m/4)}.
\label{verysmallscale}
\ee 
{}From equations (\ref{smallscale}) and (\ref{verysmallscale}) 
we see that 
$n=0$, and (now referring to equations (\ref{eq:vscale}) and 
(\ref{verysmallscale})) that
$m=0$ as well.  In other words, adopting the hypothesis that reconnection
is limited by rate of interchange of the tangle of reconnected 
flux elements that cross the
reconnection layer we conclude only that $V_{rec,global}\lesssim V_A$.
Since this limit is less stringent than 
equation (\ref{eq:lim2a}) we see that this process is not, in fact, the
bottleneck for turbulent reconnection and may therefore
be ignored.  

Our argument assumes that the scale $\Delta$ is greater than
the spectrum cutoff, $k_{\perp,max}^{-1}$, for the external turbulence
and that $\Delta'$ is similarly greater than the local spectrum
cutoff $k'_{\perp,max}$.  We can check this by adopting the {\it very}
conservative constraint that the flux element widths are greater
than the scale imposed by Ohmic diffusion with the time necessary
for a reconnection front to propagate over the reconnection layer,
that is
\be
\Delta>\left({\eta L_x\over V_{rec,global}}\right)^{1/2}\approx
\left({\eta\over L_x V_A}\right)^{1/2} L_x 
\left({V_A\over V_{rec,global}}\right)^{1/2},
\label{eq:dlim}
\ee
and
\be
\Delta'>\left({\eta \Delta\over V_{rec,global}}\right)^{1/2},
\label{eq:dlimp1}
\ee
or
\be
{\Delta'\over \Delta}>\left({\eta \over \Delta V_{rec,global}}\right)^{1/2},
\label{eq:dlimp2}
\ee
{}From equation (\ref{eq:eta}) we have
\be
k_{\perp,max}\Delta> \left({v_l L_x\over\eta}\right)^{1/4} 
\left({v_l\over V_{rec,global}}\right)^{1/4}
\left({V_A\over V_{rec,global}}\right)^{1/4}
\left({L_x\over l}\right)^{1/4}.
\ee
Since all these factors, except possibly the last, are larger than one,
and the first is usually much larger than one, we can conclude that
$k_{\perp,max}\Delta>1$.  Similarly, the local turbulence gives us
a modified limit on $k_{\perp}$ which is
\be
k'_{\perp,max}\sim \Delta^{-1} \left({V_{rec,global}\over V_A}\right)^{1/4}
\left({V_A\Delta\over\eta}\right)^{3/4}.
\ee
Combining this with equation (\ref{eq:dlimp2}) we have
\be
k'_{\perp,max}\Delta'> \left({\eta\over\Delta V_{rec,global}}\right)^{1/2}
\left({V_A\Delta\over\eta}\right)^{3/4}\left({V_{rec,global}\over V_A}\right)^{1/4}
>\left({V_A\Delta\over\eta}\right)^{1/4}\left({V_A\over V_{rec,global}}\right)^{1/4},
\ee
which is strictly greater than one.  We conclude that the our assumption of 
a self-similar hierarchy is self-consistent. 

We note that  
the fact that the horizontal transport of flux elements does {\it not}
provide the limiting constraint on $V_{rec,global}$ implies that
the fraction of the reconnection zone filled with already reconnected
flux elements, $f_r$, is not necessarily large.  However, there is firm lower
limit on $f_r$.  Since $V_{rec,global}=f_r V_{eject}$ then since
$V_{eject}\le V_A$,
\be
f_r>{V_{rec,global}\over V_A}.
\ee
If reconnection is fast then $f_r$ must be of order unity.

The third limit involves the width of the zone through which
the shared component of the magnetic field is ejected, $w_{eject}$.  
Given the stochastic geometry of the initial field configuration,
and its evolution during reconnection, this is not necessarily
the same as the width of the reconnection zone.  In any case
we have
\be
V_{rec,global} L_x= V_{eject} w_{eject}=w{V_{rec,global}\over f_r},
\ee
or
\be
w_{eject}=f_r L_x.
\ee
In a certain sense, this is just a restatement of the hypothesis that
the ejection of the shared flux from the contact volume controls the
speed of reconnection.  However, there is another connection between
$w_{eject}$ and $f_r$ where causality runs in the other direction.
If we consider the initial surface layer of flux elements facing the
reconnection zone, they have connections which run deep into the
underlying magnetized region (cf. equations (\ref{eq:diff2}) and (\ref{eq:diff3})).
However, these connections have a small volume filling factor.
Once these initial elements reconnect, further reconnection involves
flux elements that are pressed down between the reconnected elements.
This gives them an extra component normal to the plane of the 
reconnection zone, which we can express as a displacement of the
flux element from its initial position.
As $f_r\rightarrow 1$ (or at least a fraction of order unity) this
displacement is typically as large as the diffusion limit given
by equations (\ref{eq:diff2}) and (\ref{eq:diff3}) and the volume
filling factor of the reconnected flux elements rises to one.
Since typical flux elements will acquire a vertical stretch 
equal to this displacement, we see that this will also be the
width of the escape zone for the shared flux.  In other words,
if reconnected flux elements fill up the reconnection zone, then
$w_{eject}$ will be of order $f_r$ and shared flux can escape
without any difficulty.

Is this a likely scenario?  If the reconnection zone is sparsely
filled, then nearby flux elements will be forced into it by the
local pressure excess.  Since matter can escape without difficulty
in this model, this pressure excess is equal to the magnetic
pressure.  This will not be sufficient to force magnetic field
lines to bend at sharp angles, unless the whole excess is supported
on a few field lines.  However, it will be sufficient to force
nearly parallel field lines to move, so we can envision this motion
as consisting almost entirely of motion towards the reconnection zone,
without any significant bending away from the initial mean field
direction parallel to the reconnection zone.  
Furthermore, as long as $w_{eject}$ is no larger than the 
value of $k_{\perp}^{-1}$ corresponding to parallel wavelengths
$\sim L_x$, then the turbulent motions within the magnetized
fluid can be assumed to keep moving flux elements up to the
reconnection zone.  This will ensure that
the reconnection zone is continuously supplied with magnetic elements
to reconnect, and consequently the $f_r$ will grow until the loss
of magnetic flux from the reconnection zone balances the supply.
This implies that $w_{eject}$ will grow until some other process
limits the global speed of reconnection.

The fourth, and last, limit is equation (\ref{eq:lim2a}).  We see from the three
preceding arguments that this is not just an upper limit, but a reasonable
estimate for the global speed of reconnection in a turbulent medium.  
Reconnection will generally occur at a substantial fraction of the
Alfv\'en speed with the level of turbulence in the medium (or alternatively,
the level of stochasticity in the magnetic field) controlling the
exact speed.

We can turn this line of reasoning around, and use the global speed of
reconnection to estimate the properties of the reconnection zone
itself.  First, we note that the field lines will bend down to
the reconnection zone with a typical displacement equal to the
diffusion distance, or for $L_x<l$,
\be
w_{eject}=\left({v_l\over V_A}\right)^2 {L_x^{3/2}\over l^{1/2}}.
\ee
This represents a systematic bending which is just big enough to 
compete with random diffusion near the central parts of the
contact volume.  Near the edges of the reconnection zone it will
be much more conspicuous.  Also by our previous reasoning, we 
can expect that
\be
f_r\sim \left({v_l\over V_A}\right)^2\left({L_x\over l}\right)^{1/2}< 1.
\ee

The field lines will bend drastically, with a large concentration of
current and Ohmic dissipation, only in the reconnection zone itself.
If we assume that the reconnection zone and current
sheet are identical then
\be
V_{rec,local}={\eta\over\Delta}.
\label{eq:width}
\ee
This is probably not quite right, but the effect of tearing modes
appears to be only a modest enhancement of reconnection rates, judging
from the results of Appendix C, and the turbulent motions of reconnected 
field lines seem likely to disrupt modes with $k_{\|}\ll k_{\perp}$.
We should nevertheless bear in mind that equation (\ref{eq:width}) may lead to an
underestimate of the reconnection zone width, and an overestimate of
Ohmic heating within it.

Since global reconnection speeds are given by equation (\ref{eq:lim2a}),
and since the accumulation of reconnected flux elements in the reconnection
zone does not limit the pace of reconnection, we can use 
\be
V_{rec,global}\approx k_{\|}L_x V_{rec,local}\approx k_{\perp}k_{\|}L_x\eta,
\ee
where, as before, we have written the width of the reconnection zone in
terms of the equivalent wavenumber in the turbulent cascade outside
the reconnection zone.  Using equation (\ref{k_p2m}) we can rewrite this
expression and equate it to the global reconnection speed given in equation
(\ref{eq:lim2a}).  We find
\be
\Delta\approx l \left({\eta\over V_Al}\right)^{3/5}\left({V_A\over v_l}\right)^{2/5}
\left({L_x\over l}\right)^{3/10},
\label{thick}
\ee
where we have assumed that $L_x\le l$, for simplicity, and because this seems like
the most likely case.  The Ohmic dissipation rate, $\eta/\Delta^2$, is
actually less than $k_{\|}V_A$ because the pace of reconnection is set
by the escape of matter from the contact volume, rather than the
reconnection zone.  Consequently the ejection of matter from the
reconnection zone is inhibited and the corresponding ejection
velocity is less than $V_A$.  In other words, the build-up of
excess matter throughout the contact volume softens the pressure
gradient so that escaping fluid reaches an ejection velocity
$\sim V_A$ only after it has already left the reconnection zone.


One of the more striking aspects of our result of the global
reconnection speed is that it is
relatively insensitive to the actual physics of reconnection.
Equation (\ref{eq:lim2a}) only
depends on the nature of the turbulent cascade.  Although we
have reached this conclusion by invoking a particular model
for the strong turbulent cascade, we can see that any sensible model
will give qualitatively similar results.

The conclusion that reconnection is fast, even when the local
reconnection speed is slow, represents a triumph of global geometry
over the slow pace of Ohmic diffusion.  In the end, reconnection
can be fast because if we consider any particular flux element
inside the contact volume, assumed to be of order $L_x^3$,
the fraction of the flux element that actually undergoes
microscopic reconnection vanishes as the resistivity goes
to zero. 

\subsection{Reconnection for a modified spectrum of turbulence}

Our main conclusion, that reconnection is dramatically enhanced
in the presence of a stochastic field, does not depend on the particular
model of turbulence we adopt.  Our work here assumes a locally generated
turbulent spectrum which follows the scaling laws suggested by
GS95.  However, we depend only
on the scaling law for power as a function of $k_{\perp}$ and
the eddy anisotropy, that is $k_{\|}(k_{\perp})$.  The former
is supported by observations of the solar wind.  The latter
depends on the assumption of locality of nonlinear interactions
in wavenumber space and may be wrong (cf. Matthaeus et al. 1998). 
In Appendix D we discuss magnetic reconnection
for an arbitrary power-law relation between $k_{\|}$ and $k_{\bot}$ and
conclude that if $k_{\|}\sim k_{\bot}^{p}$ with $p>1/2$ then the 
reconnection rate is independent of the Lundquist number. 

Consider now the processes that can modify the spectrum of turbulence.
The turbulent spectrum may be changed due to: (a) 
a weak large scale magnetic field, (b) the injection of
energy on small scales as a result of reconnection, (c) 
large scale discontinuities resulting from 
bringing magnetic fields with different directions into direct 
contact (by definition this is a necessary part of reconnection), 
(d) the backreaction
of enhanced reconnection on the inverse cascade, and (e) the injection
of field stochasticity from some other region.

We discussed (a) and (b) above.  The first simply makes the
field more chaotic and enhances reconnection rates, 
since the effective size of $l$ is reduced.  The latter
has a much weaker effect, since it has only a slight effect
on the large scale field structure, but is likely to marginally
increase reconnection speeds.  Apropos (c), we expect that for strong 
magnetic fields brought into contact
the efficiency of the coupling between eddies at different sides
of the reconnection layer will be low. Consider, for example,
two flux tubes at $\pi/2$ angle. The eddies with $k_{\|, up}$ in the upper 
flux tube will be exciting perturbations of the order $k_{\|, up}^{-1}$
perpendicular to the magnetic field lines in the lower flux tube.
However the time scales for perturbations with $k_{\bot, low}=k_{\|,up}$
will be very different from the time scale of the eddies providing
excitation. Due to this mismatch in the time scales 
the crosstalk between eddies on opposite sides of the reconnection 
zone will be suppressed. 

The high speed of reconnection given by equation (\ref{eq:lim2a})
naturally leads to a question of self-consistency (our point (d)
above).  Is it reasonable
to take the turbulent cascade suggested in GS95
when field lines in adjacent eddies are capable of reconnecting?
It turns out that in this context, our estimate for $V_{rec,global}$
is just fast enough to be interesting.  We note that when considering the 
intersection of nearly
parallel field lines in adjacent eddies the acceleration of plasma
from the reconnection layer due to the pressure gradient
is not $k_{\|}V_A^2$, but rather $(k_{\|}^3/k_{\perp}^2)V_A^2$,
since only the energy of the component of the magnetic field 
which is not shared is available to drive the outflow.  On the
other hand, the characteristic length contraction of a given
field line due to reconnection between adjacent eddies is only
$k_{\|}/k_{\perp}^2$.  This gives an effective ejection rate 
of $k_{\|}V_A$.  Since the width of the diffusion layer over a 
length $k_{\|}^{-1}$ is just $k_{\perp}^{-1}$, we can replace 
equation (\ref{eq:lim2a}) with
\be
V_{rec,global}\approx V_A {k_{\|}\over k_{\perp}}. 
\ee
The associated reconnection rate is just
\be
\tau^{-1}_{reconnect}\sim V_A k_{\|}, 
\ee
which in GS95 is just the nonlinear cascade rate on the scale
$k_{\|}^{-1}$.  However, this result is general and does not
involve assuming that GS95 is correct.  
We will see in \S 5.1 that most of the energy
liberated in reconnection goes into motions on length
scales comparable to the dimensions of the reconnecting eddies,
so this energy release will not short circuit the energy cascade
described in GS95.  On the other hand, we can invert this argument
to see that reconnection can play an important role in 
preventing the buildup of unresolved knots in the magnetic field.
Such structures could play a major role in inhibiting the cascade
of energy to smaller scales, flattening the energy spectrum relative
to the predictions of GS95.  Our conclusion is that such structures 
will disappear as fast as they appear, supporting the notion that
they play a limited role in the dynamics of MHD turbulence.

Finally, we note that if the magnetic field structure is driven by
turbulence in another location, as when the footpoints of magnetic
arcades are stirred by turbulent motions, then we can evaluate its
effects in terms of the amplitude of field stochasticity and the
scaling of structure anisotropy with scale.  In the absence of any
particular model for this process, we note that the robust nature
of our conclusions, as shown in Appendix D, implies that under
these circumstances reconnection will be sensitive to the amplitude
of the induced field stochasticity, but not the details of the
turbulent mixing process. 

\section{Discussion}

\subsection{Energy dissipation}

The usual assumption for energy dissipation in reconnection
is that some large fraction of the energy given up by the
magnetic field, in this case $\sim \rho V_A^2 L_x^3$, 
goes into heating the electrons.  This is
not the case here.  Only a fraction, $\sim 1/(k_{\|}L_x)$ of any
flux element is annihilated by Ohmic heating within the reconnection
zone.  Over the entire course of
the reconnection event the efficiency for electron heating
is no greater than
\be
\epsilon_e\lesssim {\eta/\Delta\over V_{rec,global}}={V_{rec,local}\over V_{rec,global}}=
{1\over k_{\|}L_x}.
\ee
or, from equations (\ref{eq:lim2a}), (\ref{k_p2m}), and (\ref{thick}),
\be
\epsilon_e\lesssim \left({V_Al\over\eta}\right)^{-2/5}
\left({v_l\over v_A}\right)^{8/5}\left({l\over L_x}\right)^{4/5}.
\ee
The electron heating within the current sheet will not be uniform,
due to the presence of turbulence, the intermittent presence of
reconnected flux,
and any collective effects we have neglected here.  To the extent
that these are important they will also lower the electron
heating efficiency by broadening the reconnection layer. 

Naturally, the low value of $\epsilon_e$ is of little interest
when ion and electron temperatures are tightly coupled.
When this is not the case our model for reconnection may have
testable consequences.  As an example, we may consider
advective accretion flows (ADAFs), following the general
description given in Narayan and Yi (1995) in which advective
flows can be geometrically thick, and optically thin with
a small fraction of the dissipation going into electron heating. 
If, as expected, the magnetic pressure is comparable to the
gas pressure in these systems, then a large fraction of the
orbital energy dissipation occurs through reconnection events.
If a large fraction of this energy goes into electron heating
(cf. \cite{BL97}) then the observational arguments in favor
of ADAFs are largely invalidated. 
Our results suggest that reconnection, by itself, will not result in
channeling more than a small fraction of the energy into
electron heating \footnote{Other authors (e.g. Gruzinov 1998)
have suggested that MHD turbulence in a collisionless, highly 
magnetized environment may not terminate on small scales in ion heating,  
so that ADAFs will necessarily deposit a large fraction of
their heat into electrons.  However, Gruzinov's calculation rests on 
treating the turbulence 
as a collection of independent linear waves, which is an
unconvincing model for disturbances with a coherence time
comparable to a single wave period.}

We can also estimate the efficiency of the conversion
of magnetic field energy to high frequency Alfv\'en waves.  One
channel for the creation of such waves will be as waves
radiated from the turbulence in the reconnection zone.
The rate of energy emission is hard to estimate since the
nature of the turbulence in the reconnection zone is so
uncertain.  Plausibly the energy lost from the this zone in
the form of radiated Alfv\'en waves will be at least comparable to
the Ohmic dissipation, but it is difficult to make a more 
precise estimate.  On the other hand,  the region immediately outside the
reconnection zone will also be affected, via the constant 
interchange of reconnected flux and the subsequent relaxation
of field lines just outside the reconnection zone.  The
total energy involved at any one time will be 
$\sim \rho V_A^2 L_x^2\Delta$ and the rate of excitation of
Alfv\'en modes should be of order $k_{\|} V_{eject}=k_{\|}V_{rec,global}$.  
This process will operate for a time $\sim L_x/V_{rec,global}$, so
that the efficiency of wave energy generation will be
\be
\epsilon_{A}\sim \Delta k_{\|}\sim
\left({\eta\over V_A l}\right)^{1/5}
\left({V_A\over v_l}\right)^2 
\left({L_x\over l}\right)^{1/10}.
\ee
The associated wave frequencies will be distributed between $k_{\|}V_{rec,global}$
and $V_A/\Delta$ depending on the nature of the turbulent zone, that is
\be
{V_A\over l}\left({V_Al\over\eta}\right)^{2/5}\left({v_l\over V_A}\right)^{2/5}
\left({l\over L_x}\right)^{1/5}<\omega_A<
{V_A\over l}\left({V_Al\over\eta}\right)^{3/5}\left({v_l\over V_A}\right)^{2/5}
\left({l\over L_x}\right)^{3/10}.
\label{eq:nu1}
\ee
These limits differ by a factor of $(V_Al/\eta)^{1/5}$, which is large enough
to make precise predictions difficult, but small enough that we have a qualitative
sense of the results.

These waves are unlikely to affect of the topology of the
magnetic field, since their typical scale is, at most, only
$\lambda_{\|}$  \footnote{However, the
large scale relaxation of magnetic field lines is
likely to feed energy back into the turbulent cascade,
providing a nonlinear feedback mechanism which will
accelerate reconnection when the initial field stochasticity
is small.}.  However, they may play a role in the anisotropic
heating of ions in stellar coronae.  In this connection we note
recent observations of anisotropic velocity distributions for
HII and OVI ions in the solar corona
(\cite{KETC98}) which they suggest may be due to the presence of
high frequency (tens of kilohertz) Alfv\'en waves (see \cite{CFK98}
for a detailed analysis).

\subsection{Dynamos}

To enable sustainable dynamo action and, for example, generate 
a galactic magnetic field, it is necessary to reconnect
and rearrange magnetic flux on a scale similar to a galactic
disc thickness within roughly a galactic turnover time ($\sim 10^8$~years).
This implies that reconnection must occur at a 
substantial fraction of the Alfv\'en velocity.
The preceding arguments indicate that such reconnection velocities should
be attainable if we allow for a realistic magnetic field
structure, one that includes both random and regular fields.

One of the arguments against traditional mean-field dynamo theory 
is that the rapid
generation of small scale magnetic fields suppresses further dynamo
action (e.g. Kulsrud \& Anderson 1992). Our results thus
far show that a random magnetic field enhances reconnection by
enabling more efficient diffusion of matter from the reconnection
layer. This suggests that the existence of small scale magnetic
turbulence is a prerequisite for a successful large scale dynamo.
In other words, we are arguing for the existence of a kind of 
negative feed-back. If the magnetic field is too smooth, reconnection
speeds decrease and the field becomes more tangled.  If the field
is extremely chaotic, reconnection speeds increase, making the field
smoother. We note that
it is common knowledge that magnetic reconnection can sometimes be
quick and sometimes be  slow. For instance, the existence bundles of flux
tubes of opposite polarity in the solar convection zone indicates that 
reconnection can be very slow. At the same time, solar flaring suggests 
very rapid reconnection rates.    

There is a general belief that magnetic dynamos operate in stars,
galaxies (see Parker 1979) and accretion disks (see Balbus \& Hawley 1998).
In stars, and in many accretion disks, the plasma has a high $\beta$, 
that is the average plasma pressure is 
higher than the average magnetic pressure. In such situations 
the high diffusivity of magnetic field can be explained
by concentrating flux in tubes (Vishniac 1995a,b). This trick does not
work in the disks of galaxies, where the magnetic field is 
mostly diffuse (compare Subramanian 1998)
and ambipolar diffusion impedes the formation of flux tubes (Lazarian
\& Vishniac 1996).  
This is the situation where our current treatment
of magnetic reconnection is most relevant.  However, 
our results suggest that magnetic reconnection proceeds regardless
and that the concentration of magnetic flux in flux tubes
via turbulent pumping is not a necessary requirement for 
successful dynamos stars and accretion discs.  

We do not address here the controversial issue of the turbulent
dynamo in clusters of
galaxies. This was first suggested by (Jaffe 1980) and was elaborated 
in great
detail by Ruzmaikin, Sokoloff \& Shukurov (1989), who claimed
an excellent match between observations and predictions based
on the  Kazantsev (1968) theory of the turbulent dynamo.
However, Goldshmidt \& Rephaeli (1993)
found a large ($\sim 10^{20}$) numerical error
in the value of Ohmic diffusivity used by Ruzmaikin, Sokoloff \&
Shukurov (1989), which formally invalidated their result.
However, if it were possible to use the effective diffusivity
determined by the reconnection rate instead of Ohmic diffusivity,
then the theory of turbulent dynamo can be revived for clusters of galaxies.
We postpone the discussion of this important question until our next paper. 

Our results show that in the presence of MHD turbulence magnetic
reconnection is fast, and this in turn allows the possibility
of `fast' dynamos in astrophysics (see the discussion of the {\it fast
dynamo} in Parker (1992)). As the fluid
conductivity tends to zero the relevant parallel wavelength also
goes to zero, allowing a 
change of magnetic field topology via reconnection at a finite rate.

\subsection{Turbulent diffusion, particle diffusion, and self-excited reconnection}

It may seem from our discussion that this paper revives the 
concept of turbulent diffusion. However, a closer comparison between
the concept original concept of turbulent diffusivity and our
suggestion shows that the two are very different.
Within the turbulent diffusivity paradigm, magnetic fields of different
polarity were
believed to filament and intermix on very small scales while 
reconnection was believed to proceed slowly.  On the contrary,
we have shown that the global speed of reconnection is fast if 
a moderate degree of magnetic field line wandering is allowed. 
The latter, unlike the former,
corresponds to a realistic picture of MHD turbulence and does not
entail prohibitively high magnetic field energies at small
scales.

On the other hand, the diffusion of particles through a magnetized
plasma is greatly enhanced when the field is mildly stochastic.
There is an analogy between the reconnection problem and the diffusion
of cosmic rays (Barghouty \& Jokipii 1996). In both cases charged 
particles follow
magnetic field lines and in both cases the wandering of the magnetic field
lines leads to efficient diffusion.

Finally, we note that observations of Solar flaring 
seem to show that such events start from some limited
volume and spread as though
a chain reaction from the initial reconnection region initiated 
a dramatic change in the magnetic field properties. Indeed, 
Solar flaring happens as if the
resistivity of plasma were increasing dramatically as plasma 
turbulence grows (see Dere 1996 and references therein).  
In our picture this is a consequence of
the increased stochasticity of the field lines rather than
any change in the local resistivity.  The change in magnetic
field topology that follows localized reconnection provides   
the energy necessary to feed a turbulent cascade in neighboring
regions.  This kind of nonlinear feedback is also seen in the
geomagnetic tail, where it has prompted the suggestion that
reconnection is mediated by some kind of nonlinear instability
built around modes with very small $k_{\|}$ (cf. Chang 1998 and
references therein).  The most detailed exploration
of nonlinear feedback can be found in the work of Matthaeus and
Lamkin (1986), who showed that instabilities in narrow current
sheets can sustain broadband turbulence in two dimensional
simulations.
Although our model is quite different, and relies on the
three dimensional wandering of field lines to sustain fast
reconnection,  we note that the concept
of a self-excited disturbance does carry over and may describe
the evolution of reconnection between volumes with initially
smooth magnetic fields.

\section{Conclusions}

In this paper we have explored the consequences of a small stochastic
component for the reconnection of large scale magnetic fields.
We have used a particular model of MHD turbulence, due to Goldreich
and Sridhar (1995), in order to draw firm conclusions, but the
main features of our results should follow from any model
which is consistent with our observations of MHD turbulence in
astrophysical systems.  We have implicitly assumed that viscosity
is less important than resistivity in setting the limits for
any turbulent cascade.  Although this should usually be the case
for astrophysical plasmas, it is not guaranteed, and the effects of
viscosity will need to be considered in more detail 
(Lazarian \& Vishniac
1999).  Finally, we have
assumed that we are dealing with a strong magnetic field, where motions
that tend to mix field lines of different orientations are
largely suppressed.  The galactic magnetic field is usually taken
to have grown via dynamo action from some extremely weak seed field
(cf. Zel'dovich, Ruzmaikin \& Sokoloff 1983, Lazarian 1992, Kulsrud, Cen,
Ostriker, \& Ryu 1997 and references contained therein).  
When the field is weak it can be moved as a passive
scalar and its spectrum will mimic that of Kolmogorov turbulence.
The difference between $\lambda_{\bot}$ and $\lambda_{\|}$ vanishes,
the field becomes tangled on small scales, and 
$V_{rec, local}$ becomes of the order of $V_A$.
Of course, in this stage of  evolution $V_A$ may be very small.
However on the small scales $V_A$ will grow to equipartition
with turbulent velocities
on the turnover time of the small eddies. The enhancement of reconnection
as $V_A$ increases accelerates the inverse cascade as small magnetic
loops merge to form larger ones.

We list the principal results of this paper below.

First, the rate of magnetic reconnection is increased dramatically 
in the presence of a stochastic component to the magnetic field. 
This component arises naturally whenever turbulence is present.
Even when the turbulent cascade is weak the resulting reconnection
speed is independent of the Ohmic resistivity. 

Second, the argument that the rapid rise of random magnetic field associated
with dynamo action results in the  suppression of dynamo (Kulsrud \&
Anderson 1992) is untenable since the increase of the random component
of the magnetic field increases the reconnection rate. We conclude that 
dynamo is a self-regulating process.

Third, the second parameter in determining the reconnection 
speed is not some aspect of the microphysics, but the level of
field stochasticity (or the large scale kinetic energy feeding
the turbulent cascade).  As reconnection proceeds the local turbulent
cascade will grow stronger and the initial level of stochasticity
will matter less and less. On the contrary, microphysical
processes widely believed to speed up reconnection, 
i.e. anomalous resistivity, fail in interstellar conditions.

Fourth, since the speed of reconnection depends on the presence of
small scale turbulence, non-resistive processes which terminate the 
turbulent cascade may slow reconnection down.  As an example, we
can cite neutral-ion collisions in a dense, partially ionized gas. 
On the other hand, the rate of collisions in a fully ionized
plasma is irrelevant.  Under typical astrophysical conditions
$\nu\ll\eta$ for fully ionized plasmas.  When collisions become
very rare, the only effect is to eliminate the cascade of compressive
modes, which are not important for reconnection.

Fifth, reconnection in a turbulent medium deposits only a small
fraction of the magnetic field energy directly into heating electrons. 
Whether or not this is important depends on the rate of energy
exchange between ions and electrons, and the unknown degree
to which the turbulent energy cascade can avoid depositing its
energy into electrons.  

Sixth, regardless of the other arguments advanced here, 
there is a minimal reconnection speed, 
$\sim V_A {\cal R}_L^{-3/16}$, where ${\cal R}_L\equiv (V_Al/\eta)$,
which is much faster than the Sweet-Parker estimate, even though
it is still unrealistically slow.

Finally, the current sheet associated with reconnection is not 
substantially broadened from the Sweet-Parker
result, even when reconnection is fast.

\acknowledgements

We acknowledge some helpful suggestions on the presentation of this 
material by the referee, Eugene Parker.
We are thankful to B. Draine for 
helpful discussions and to Phil Kronberg for a critical 
reading of the manuscript. Elucidating comments by Bill Matthaeus and
George Field are acknowledged. This work was supported in
part by  NASA grants NAG5-2858, NAG5-7030 (AL), NAG5-2773 (ETV), and
NSF grant AST-9318185 (ETV).  The final stages of this work were supported
for one of us (AL) by CITA Senior Research Fellowship. ETV
is grateful for the hospitality of MIT and
the CfA during the early stages of this work, and CITA during
its conclusion. 

\appendix

\section{The Power Spectrum of Magnetohydrodynamic Turbulence}

Here we discuss a model for MHD turbulence first proposed in GS95.
Although their work included an explicit calculation of 
coupling constants based on treating the turbulent motions
as shear Alfv\'en waves, the important parts of the model
depend only on some rather general scaling arguments, which
we will repeat here.  These arguments depend on assuming
that power is transferred locally in Fourier space, which
is a convenient, but not rigorously justified, assumption.

We begin by noting that the first simple model of incompressible 
MHD turbulence was proposed independently by Iroshnikov (1963) and
Kraichnan (1965) based on the
interactions of triads of waves (see also Diamond and Craddock 1990).
In this picture the energy transfer rate is roughly
\be
\tau_{nl}^{-1}\approx {(k v_k)^2\over \omega_A}
\label{tau}
\ee 
where $k$ is the magnitude of a wavevector, $v_k$ is the rms fluid velocity 
contributed from power on the scale $k^{-1}$, $\omega_A\equiv k_{\|}V_A$ is 
the Alfv\'en wave frequency, $k_{\|}$ is the
wavevector component parallel to the magnetic field direction, and $V_A$
is the Alfv\'en velocity.  In the original picture the power was assumed
to spread isotropically in wavevector space.  As long as $v_k<V_A$
this expression for the nonlinear time scale will be less than $kv_k$,
that is, the magnetic field reduces the cascade of energy to
higher wavenumbers.

In a recent series of papers Goldreich and Sridhar 
(Sridhar and Goldreich 1994; Goldreich and Sridhar 1995; Goldreich and
Sridhar 1997) questioned the validity of this picture
and argued that the diffusion of power towards larger
values of $k_{\|}$ is strongly suppressed.  Their original
claim that three wave interactions are completely suppressed has
been strongly criticized (Montgomery and Matthaeus 1995,
Ng \& Bhattacharjee 1996).  However, in their most recent paper
(Goldreich and Sridhar 1997) they showed convincingly  
that the effect of residual
three wave couplings is consistent with a picture in which  
the basic nonlinear time scale is set by equation (\ref{tau})
but with an anisotropic spectrum in which virtually all of the
transfer of power between modes moves energy towards larger
$k_{\perp}$ while leaving $k_{\|}$ unchanged.  They proposed
calling this regime `intermediate turbulence',
since while the nonlinear decay rate is identical to the
usual expression for weak turbulence among dispersive waves,
in this case the higher order mode couplings are all comparably
important.  If we invoke the constancy of the local energy
flux through the cascade, $v_k^2/\tau_{nl}$, as a function of scale, 
then from equation (\ref{tau}) we see that in this regime
\be
v_k\propto k_{\perp}^{-1/2},
\label{eq:vk2}
\ee
where we have assumed that $k_{\perp}>>k_{\|}$.

As the power cascades to larger values of $k_{\perp}$
the magnetic field becomes progressively less important
in the mode dynamics.  Eventually we have
\be
k_{\|}V_A\le kv_k.
\label{eq:parcrit}
\ee
In this limit the motions are no longer wave-like, the magnetic
field exerts only a weak influence on the dynamics, and the fluid motions
resemble ordinary hydrodynamical turbulence with a nonlinear
time scale $\sim kv_k$.  Since $k_{\|}$
is no longer privileged, the cascade of power is in the direction of
increasing isotropy. Since $k_{\perp}\gg k_{\|}$ this implies an 
increase in $k_{\|}$.

If energy is injected on some scale $l$, with $v_l\le V_A$, then we expect
the cascade to transfer energy to larger $k_{\perp}$ until the condition
expressed in equation (\ref{eq:parcrit}) is satisfied.  At this point
the turbulence is no longer wavelike (since $\tau_{nl}^{-1}\sim k_{\|}V_A$).
However, since turbulence tends towards isotropy when the magnetic field
is completely negligible, we expect $k_{\|}$ and $k_{\perp}$
to increase in tandem so that equation (\ref{eq:parcrit}) is just
marginally satisfied.  This is the regime of strong turbulence
described in GS95.  At all smaller parallel 
wavelengths fluid motions bend magnetic field lines easily.
Consequently we expect most of the power in energy spectrum to be
centered around wavenumbers such that 
\be
k_{\|}\sim k_{\perp}{v_k\over V_A}
\label{k_p}
\ee
We approximate the energy transfer rate for the turbulent 
cascade,  $\dot{\cal E}$, for $v_l\le V_A$ as
\be
\dot {\cal E}\approx {v_l^4\over l V_A},
\label{eq:casc}
\ee
if  $v_l\le V_A$ on the scale $l$.
The usual hydrodynamic choice
$\dot {\cal E}\approx {v_l^3\over l}$ is valid otherwise,
although not relevant for our present discussion.  When
the magnetic field is weak and the largest scales in the turbulent cascade
are essentially hydrodynamic then we can identify $l$ with the
scale of equipartition so that $v_l^2\sim V_A^2$. 

Conservation of energy in the turbulent cascade implies that
$\dot{\cal E}=v_k^2/\tau_{nl}=const$ and is only determined by
large scale motions. If $l$ is the large scale eddy size, and
$v_l<V_A$ then we use equation (\ref{eq:casc}) to find $\dot{\cal E}$. 
{}From equation (\ref{tau}) we find that
\be
\dot{\cal E}={v_l^4\over l V_A}\sim {v_k^2\over\tau_{nl}}\sim v_k^4
{k_{\perp}^2\over k_{\|}V_A}.
\label{cass}
\ee
Combining equations (\ref{k_p}) and (\ref{cass}) we obtain
\be
k_{\|} \approx l^{-1} (k_{\perp}l)^{2/3}
\left({v_l\over V_A}\right)^{4/3}.
\label{k_p2}
\ee
Substituting this in equation (\ref{tau}) gives the energy transfer rate 
\be
\tau_{nl}^{-1}\approx {v_l\over l}(k_{\perp}l)^{2/3}\left({v_l\over V_A}\right)^{1/3},
\label{tau2}
\ee
while the rms fluid velocity on the scale $k$ follows from equation 
(\ref{k_p}) 
\be
v_k\approx v_l (k_{\perp}l)^{-1/3}\left({v_l\over V_A}\right)^{1/3}.
\label{vel}
\ee

Had we assumed an isotropic distribution of power in $\vec k$ space
(that is, had we failed to distinguish between $k_{\perp}$ and
$k_{\|}$)
we would have recovered the Kraichnan spectrum, with $v_k\propto k^{-1/4}$.
Instead, we find $v_k\propto k_{\|}^{-1/2}$.
Curiously enough, this anisotropic cascade looks qualitatively like 
the Kolmogorov spectrum when expressed in terms of $k_{\perp}$.
This becomes less surprising
if one recalls that the magnetic field does not influence motions that
do not bend it.  We also note that for the characteristic $k_{\|}$
given in equation (\ref{k_p2}) the interaction between different modes 
can be viewed, with equal validity, as a wave-wave interaction, an 
eddy-eddy interaction, or a wave-eddy interaction.  That the first two
give roughly equal nonlinear time scales is immediately apparent from
equations (\ref{tau}), and (\ref{k_p}).  That the
third point of view is also equivalent follows from the work of
Similon and Sudan (1989) who showed that the dissipation of Alfv\'en
waves in a stochastic field can be estimated by replacing the microscopic
Ohmic resistivity by the diffusion coefficient for waves traveling along
the field lines.  In terms of the turbulent cascade envisioned here, this
is equivalent to making the substitution
\be
\eta\rightarrow \eta_{turb}\approx \Delta^{2}\omega_A,
\ee
where $\Delta$ is the typical scale of perpendicular displacements of the field
lines, and also the correlation length for such displacements.  In our
model, for typical modes, $k_{\perp}\Delta\sim 1$.
It follows that $k_{\perp}^2\eta_{turb}$ is equal to other estimates of the
nonlinear interaction time scale.

The maximum $k_{\perp}$ can be obtained from the condition that the energy
transfer rate (see eq.~(\ref{tau2})) be greater than the resistive 
dissipation rate $k_{\bot}^2 \eta$. We get
\be
k_{\perp}l<k_{\perp,c}l\approx
 {v_l\over V_A}\left({V_A l\over \eta}\right)^{3/4}.
\label{eq:eta}
\ee
Analogously, viscosity will be negligible as long as $\tau_{nl}^{-1}$ is
greater than the viscous dissipation rate $k_{\bot}^2 \nu_{\bot}$, which
implies that
\be
k_{\perp}l< {v_l\over V_A}\left({V_A l\over \nu_{\bot}}\right)^{3/4},
\ee
where we have written the viscous diffusivity as $\nu_{\bot}$ to emphasize that
it is the viscosity perpendicular to magnetic field lines which 
enters into this estimate.
{}For a pure plasma the viscous diffusivity may be large when 
the density and magnetic field are both low, but
it will decrease as the field strength increases. 

In partially ionized
gas, collisions with neutrals provide an additional source of viscosity. 
In this case the ambipolar diffusion
coefficient $\eta_{amb}$ should be used instead of $\eta$ in 
Eq.~(\ref{eq:eta}). For typical conditions in the cold interstellar
medium (see Vishniac \& Lazarian 1998)
\be
\eta_{amb}\approx 5.3\times 10^{21}n^{-1}_{0.03}N_{30}^{-1}T^{-1/2}_{100}
B^2_{5}~{\rm cm}^2/{\rm s}~~,
\label{eq:ambi}
\ee
where the letters with 
subscripts denote the normalized values of the parameters
adopted. In other words, the electron density $n$ is normalized on the
density $0.03$~cm$^{-3}$, neutral density $N$ is normalized by 
$30$~cm$^{-3}$, typical temperature $T$ is assumed 100~K, while
typical magnetic field $5\times 10^{-6}$~Gauss. Such a high
diffusivity implies $k_{max}l$ is only $10^3$ for $l=30$~pc, so that
turbulence will be damped at $\sim 0.03$~pc scales in the
cold interstellar medium.

Rather high values of diffusivity follow from collisionless damping
(see Foote \& Kulsrud 1969), but this only affects compressional modes
in the cascade. GS95 showed that very little energy is transferred
to the compressional modes from the shear modes on scales for which
$k_{\perp}\gg k_{\|}$. Consequently, the suppression of compressional
modes has little direct impact on the small scale structure of the
magnetic field.

The minimum $k_{\perp}$ covered by these scaling laws follows from equation
(\ref{k_p2}).  We find that
\be
k_{\perp}l> \left({V_A\over v_l}\right)^{2}.
\ee
Larger perpendicular scales will be described by weak Alfvenic
turbulence.

Are these scaling laws consistent with observations?
The Sun provides our best example of strong MHD turbulence, but current
observations of the solar corona and photosphere do not significantly 
constrain the power spectrum of the inertial range.  On the other hand, the
solar wind contains dynamical disturbances on a broad range of scales.
Barnes (1979) has advocated that these should be interpreted
as Alfv\'en waves emitted from the solar corona and undergoing only
weak dissipation in their outward journey.  However
Coleman (1968) has argued that the shorter wavelength disturbances
are most naturally interpreted as a turbulent spectrum, and the
solar wind shows evidence for local dissipation on small scales.
A recent analysis by Leamon et al. (1998) supports the latter
interpretation and argues that between $1.6\times10^{-4}$ and
$4\times 10^{-2}$ Hz perturbations in the solar wind are in the
inertial range of strong MHD turbulence.  (Quoted frequencies are
measured at a spacecraft essentially stationary in a moving solar wind.)
The power spectrum in this range has an index of roughly $-0.7$, 
consistent with a Kolmogorov index of $-2/3$.  Since the solar wind
advects disturbances over the spacecraft radially, while the 
unperturbed magnetic field vector is strongly tilted by the Sun's
rotation, this is essentially a measurement of the scaling law
with respect to $k_{\perp}$.  The observations also indicate that
the turbulence is strongly anisotropic.  This is in qualitative
agreement with GS95, although a quantitative comparison
would require an estimate of the fraction of the energy contained
in nonlinear magnetosonic waves, and the local value of $P_{mag}/P_{gas}$.
This raises questions that have little impact on the nature of the
small scale stochastic magnetic field structure, and we will not 
pursue them further.

MHD turbulence in the ISM can be studied via scintillations. 
These measurements
(see Spangler 1998) reveal that the density irregularities responsible 
for radio wave scintillation are elongated and aligned along the 
magnetic field.
The scintillations themselves are believed to result from entropy 
fluctuations
that arise from the mixing of plasma elements having different 
specific entropies.
The entropy acts as a passive scalar and consequently 
its spectrum assumes the 
form of the turbulent energy spectrum (see Lesieur 1990). As the
sharpest entropy gradients lie in the plane perpendicular
to the magnetic field, we would expect to see the spectrum $k^{-5/3}$.
It is encouraging that the measurements (see Armstrong, Rickett \& Spangler
1997) do get such a spectrum. Moreover, the elongation of fluctuations
along the field lines follows from the disparity between $k_{\|}$
and $k_{\bot}$. This may also serve as a hint in favor of Goldreich
and Sridhar's model\footnote{The range of scales over which the spectrum
has this form is hotly debated. The emissivity of
HI over the scale 10-100 pc has a shallower spectrum in the
velocity space defined by galactic rotation (Lazarian 1995). 
Lazarian \& Pogosyan (1999) claim that this spectrum is consistent with
the Goldreich \& Sridhar's model of turbulence.
However, here we are mostly interested in the smallest
scales.}.


Numerical simulations also seems to be consistent with the GS95 model. The high wavenumber
resistive cutoff is given by Eq.~(\ref{eq:eta}). Together with Eq.~(\ref{k_p2})
this sets the upper limit to $k_{\|}$:
\be
k_{\|,c}\approx l^{-1} \left({v_l\over V_A}\right)^2\left({lV_A\over\eta}\right)^{1/2}.
\ee
We immediately see that, if $b$ is the perturbation of magnetic field
at the scale $l$,
\be
{k_{\|,c}\over k_{\perp,c}}\approx {v_l\over V_A} \left({\eta\over lV_A}\right)^{1/4}\approx {b\over B} \left({\eta\over lV_A}\right)^{1/4}~~~,
\ee
which is similar to the scaling 
relation\footnote{Strictly speaking,
in Matthaeus et al (1998) the relation $k_{\|,c}/ k\propto
b/B$, where
$k=(k_{\|,c}^2+ k_{\perp,c}^2)^{1/2}$, was reported. The scaling above
differs
from this by $\sim (b/B)^{1/4}$, which for the limited dynamical range 
available provides a difference of the order 30\%. However,
the difference may not be statistically significant. Peter Goldreich
(private communication) attracted our attention to the fact that
in Matthaeus et al (1998) $k_{\|}$ is defined solely in terms of
the background field and therefore if $k_{\|}$ is smaller than
$k_{\bot}\theta$, where $\theta$ is the typical large scale bending angle,
the local value of $k_{\|}$ is not measured by this statistic. This 
will bias the measurements of the anisotropy towards a linear
relationship. Definitely, new, more elaborate numerical
tests are needed!}    reported in Matthaeus et al (1998).

\section{Previously explored ways to speed up reconnection}

\subsection{Anomalous resistivity}

Anomalous resistivity is present in the reconnection  layer when the
field gradient is so sharp that the electron drift velocity is 
of the order of thermal velocity of ions $u=(kT/m)^{1/2}$.
For a detailed discussion of the physics see Parker (1979).
The condition for appearance of anomalous resistivity is
\be
j>j_{cr}=Neu~~~.
\label{current}
\ee
If the current sheet has a width $\delta$ with a change in the
magnetic field $\Delta B$ then 
\be
4\pi j=\frac{c \Delta B}{\delta}~~~.
\label{dense}
\ee
The effective resistivity increases nonlinearly as $j$ becomes greater
than $j_{cr}$, thereby broadening the current sheet.
We can therefore assume that $j$ stays of the order of $j_{cr}$,
that is
\be
\delta\approx\frac{c\Delta B}{4\pi N e u}~~~.
\label{deltt}
\ee
Expressing $\delta$ through ion cyclotron radius $r_c=(mu c)/(e B_{tot})$,
where $B_{tot}$ is the total magnetic field (including any shared
component) one gets
\be
\delta\approx r_c \left(\frac{V_A}{u}\right)^2 \frac{\Delta B}{B_{tot}}~~~,
\label{deltrc}
\ee
which agrees with Parker (1979) up to the factor
$\Delta B/B_{tot}$, which is equal 1 in Parker's treatment.

We can derive a reconnection speed by combining 
mass conservation and equation (\ref{deltrc}). We get
\be
V_{rec}\approx V_A \frac{r_c}{L_x} 
\left(\frac{V_A}{u}\right)^2 \frac{\Delta B}{B_{tot}}~~~.
\label{rec:anom}
\ee

Equation (\ref{rec:anom}) shows that the enhanced
reconnection velocity is much less than
the Alfven velocity if the characteristic scale of the problem
$L_x$ is much greater than the ion Larmor (cyclotron) radius.
In general, ``anomalous reconnection'' is important when
the thickness of the reconnection layer in the Sweet-Parker
reconnection scheme is less than $\delta$. This means that when
\be
\delta>\left(\frac{L_x \eta}{V_A}\right)^{1/2}~~~,
\label{unequal}
\ee
anomalous effects are important. The condition (\ref{unequal})
is easily fulfilled in the laboratory, where anomalous effects
are widely observed. However, for typical interstellar magnetic
fields the Larmor radius $r_c$ is $\sim 10^7$~cm. For 
interstellar Ohmic diffusivities $\eta\approx 10^9$~cm$^2$/sec
and $V_A\approx u\sim 10^6$ cm/sec we find that anomalous resistivity
becomes important if
\be
L_x<\frac{\delta^2 V_A}{\eta}\approx 10^{11}~{\rm cm}\approx 0.01 {\rm AU}~~~,
\label{Lanom}
\ee
which is a very small scale for the ISM. All in all,
anomalous effects are unlikely to be important for interstellar 
reconnection if the Sweet-Parker scheme is involved.

In the presence of magnetic field turbulence the exact value
of resistivity is much less significant.  Nevertheless,
anomalous resistivity may be important within the reconnection
zone itself.  By comparison with equation  (\ref{thick}).
we can see that anomalous effects will be important if 
\be
r_c>l\left({\eta\over V_Al}\right)^{3/5}\left(V_A\over v_l\right)^{2/5}
\left(L_x\over l\right)^{3/10},
\ee 
or assuming $V_A\sim v_l$ and $L_x\sim l$
\be
l<10^{13}~{\rm cm}\approx 1{\rm AU}~~~.
\label{l_anom}
\ee 
Unfortunately, under normal conditions this is still a very small length scale.
If $l$ is less than $\sim 1$~AU the speed of reconnection can be found
from the condition $V_{rec, local} \lambda_{\|}\approx V_A \delta$, where
$\lambda_{\|}\sim k^{-1}_{\|}$ is given by equation~(\ref{k_p2}) and we assume 
$k_{\bot}$ that enters this formulae $\sim \delta^{-1}$. Substituting
$\delta$ from equation~(\ref{deltrc}) we get
\be
V_{rec, local}\approx V_A \left(r_c\over l\right)^{1/3}
\left(\Delta B\over B_{tot}\right)^{1/3}\left(V_A\over u \right)^{2/3}
\left(v_l\over V_A\right)^{4/3}
\ee

Similarly, in the presence of
the ambipolar diffusion in the gas with low degree of ionization
the thickness of the reconnection layer becomes just barely greater
than the ion Larmor radius (Vishniac \& Lazarian 1998) so that we
might expect the onset of anomalous effects, but even if they were
marginally important they would not change the reconnection velocity
significantly.  In general we see that anomalous
effects are of limited importance for reconnection in the interstellar 
medium.

Recently this topic was revisited by Biskamp et al. (1997) and
Shay et al. (1998) who showed that in a collisionless plasma
the minimum width for the ion flow is actually the ion skin
depth $c/\omega_{pi}$ rather than the ion cyclotron radius.
Since the former is just $V_A/u_i$ times the latter, this does not
affect the conclusion that anomalous effects cannot explain
fast reconnection in astrophysical plasmas \footnote{Shay et al.
also found that the reconnection speed in their simulations was
independent of $L_x$, which would suggest that something like
Petschek reconnection emerges in the collisionless regime.
However, their dynamic range was small and the ion
ejection velocity increased with $L_x$, with maximum speeds
approaching $V_A$ for their largest values of $L_x$.  Assuming that
$V_A$ is an upper limit on ion ejection speeds we may expect
a qualitative change in the scaling behavior of their simulations
at slightly larger values of $L_x$.}.

\subsection{Bohm diffusion}

Bohm diffusion is phrase which describes a process which is 
ubiquitous in laboratory plasmas, but which lacks
a good theoretical explanation. Its characteristic
feature is that ions appear to scatter about once per Larmor
precession period. The resulting particle diffusion destroys the
`frozen-in' condition and allows significant larger magnetic field
line diffusion.  For an ion thermal velocity $u\approx (kT/m)^{1/2}$
and a cyclotron radius $r_c=muc/qB$, where $q$ is the ion charge,
the Bohm particle diffusion coefficient is 
$\eta_{\rm B}\approx u r_c\approx kT (c/qB)$. If we
are concerned instead with the diffusion of field lines, we need
to substitute $V_A^2$ for $kT/m$ so that $\eta_{\rm Bohm}\sim V_A r_c$.
Since $V_A$ is of order $u$ in the ISM this has no effect on 
our estimates.  In the presence of Bohm diffusion, this coefficient 
should be used in place of Ohmic diffusivity 
(Parker 1979).  However, we note that even if we make this substitution,
it can produce fast reconnection, of order $V_A$, only if $r_c\sim L_x$.
It therefore fails as an explanation for fast reconnection for the
same reason that anomalous resistivity does. The difference between the
two processes is that anomalous resistivity vanishes when the scales
involved become larger than the threshold values given by 
equations (\ref{Lanom}) and (\ref{l_anom}), 
while we cannot define such a threshold for Bohm diffusion.

We may repeat all our calculations in the presence of Bohm diffusion
by substituting $\eta_{\rm Bohm}$ instead of $\eta$ 
and as $\eta_{\rm Bohm}\gg \eta$, the lower limit for the
reconnection rates will increase
by a substantial factor  $(\nu_{\rm Bohm}/\eta)^{3/16}\gg 1$. 

A major shortcoming of this idea is that it is not at all clear that
Bohm diffusion is as omnipresent under astrophysical 
conditions as in laboratory plasmas. Moreover, since our present study 
suggests that the actual reconnection rate does not depend on the 
value of diffusivity, this substitution is probably irrelevant.

\section{Tearing modes}

Tearing modes are a robust instability connected to the 
appearance of narrow current sheets (\cite{FKR63}).  
The resulting turbulence
will broaden the reconnection layer and enhance the reconnection
speed.  Here we give an estimate of this effect and show that
while it represents a significant enhancement of Sweet-Parker
reconnection of laminar fields, it changes our results by only
a small amount.
One difficulty with many earlier
studies of reconnection in the presence the tearing modes
stemmed from the idealized two dimensional geometry
assumed for reconnection (\cite{FKR63}). In two dimensions tearing 
modes evolve via
a stagnating non-linear stage related to the formation of magnetic
islands.  This leads to a turbulent reconnection zone (\cite{ML85}), 
but the current sheet remains narrow and its effects on the 
overall reconnection speed are unclear.
This nonlinear stagnation stage does not emerge when realistic three dimensional
configurations are considered. Indeed, it is easy to see that
instead of islands one finds nonlinear
Alfv\'en waves in three dimensional reconnection
layers.  Therefore the tearing instability proceeds at high rates 
determined by its linear growth while the resulting magnetic structures
propagate out of the reconnection region at the Alfv\'en speed.

The dominant mode will be the longest wavelength mode,
whose growth rate will be
\be
\gamma\approx {\eta\over\Delta^2}\left({V_A\lambda_{\|}\over\eta}\right)^{2/5}.
\label{eq:growth}
\ee
The transverse spreading of the plasma in the reconnection layer will 
start to stabilize\footnote{As a consequence of stabilization, tearing
mode reconnection is stationary in a time-averaged sense.} this 
mode when its growth
rate is comparable to the transverse shear $V_A/\lambda_{\|}$ (\cite{BSS79}). 
At this point we have
\be
V_{rec,local}\approx\gamma\Delta\approx V_A{\Delta\over\lambda_{\|}}.
\label{eq:shear}
\ee
For a laminar magnetic field we have $L_x=\lambda_{\|}$ and these
two equations imply
\be
V_{rec}=V_A\left({\eta\over V_A L_x}\right)^{3/10}.
\ee
This is substantially faster than the usual Sweet-Parker formula,
but still quite slow. We note in passing that the value of
$\Delta/\lambda_{\|}$ we assume here is below the minimum value
proposed by Furth et al. (1963).  However, it is consistent with
the more accurate calculations of Van Hoven and Cross (1971). 

If we consider instead a stochastic field with $\lambda_{\|}\ll L_x$,
then we can express the local
reconnection rate in terms of $k_{\|}\sim \lambda_{\|}^{-1}$. Substituting
$ k_{\|}$ from equation (\ref{k_p2})  and assuming $k_{\bot} \sim \Delta^{-1}$ we get 
\be
V_{rec,local}\approx V_A \left({v_l\over V_A}\right)^{3/4}
\left({\eta\over V_Al}\right)^{3/16},
\label{eq:vrec2}
\ee
while
\be
k_{\|}l\sim \left({v_l\over V_A}\right)^{5/2}
\left({V_Al\over\eta}\right)^{3/8},
\label{lpara}
\ee
and
\be
k_{\perp}l\sim \left({v_l\over V_A}\right)^{7/4}\left({V_Al\over\eta}\right)^{9/16}.
\label{eq:scalep}
\ee

The tearing instability broadens the reconnection region.  
For a completely ionized gas $k_{\perp}<k_{\perp,max}$, and equation
(\ref{eq:vrec2}) holds, provided that
\be
\frac{V_A l}{\eta}>\left(v_l\over V_A\right)^4~~~,
\label{eq:uneq}
\ee
which is nearly always true. 
In a partially ionized gas the tearing modes will be partly
suppressed, and this mechanism for increasing $V_{rec,local}$
will be relatively ineffective.

Equation (\ref{eq:vrec2}) is the speed of reconnection for 
each flux element in a fully ionized gas, using Goldreich and
Sridhar's model for the MHD turbulent cascade, and assuming the 
Sweet-Parker topology for the reconnection sheet.
Equations (\ref{eq:vrec1})
and (\ref{eq:vrec2}) are not dramatically different.  The exponent
of the Lundquist number, ${\cal R}_L\equiv (V_A l/\eta)$ 
changes by only $1/16$.
We also note that the frequency of
reconnection, in this case the tearing mode growth rate, is
$\sim V_A/\lambda_{\|}$, so once more we can treat the field line
structure as essentially static.

\section{Turbulence with an arbitrary spectrum}

Although in the main body of the paper we presented arguments
in favor of Goldreich \& Sridhar's (1995) model of turbulence, it is 
worthwhile considering reconnection when the relation between $k_{\|}$
and $k_{\bot}$ is of the form
\be
k_{\|} \approx l^{-1} (k_{\perp}l)^{p}
\left({v_l\over V_A}\right)^{m},
\label{k_p2a}
\ee
where $p$ and $m$ are unspecified positive constants.
For GS $p=2/3$, and $m=4/3$.  On the other hand, Matthaeus et al.
(1998) have criticized the model of Goldreich and Sridhar,
partly on the basis of numerical simulations, and proposed
an alternative model in which $p=m=1$\footnote{Note, however,
that the scaling of eddy anisotropy seen in their simulations
is in agreement with GS95, once allowance is made for the existence
of weak turbulence on large scales (cf. the discussion at the
end of Appendix A of this paper).}.  Using
equation (\ref{k_p2a})
instead of eq.~(\ref{eq:diffuse}), we get 
\be
\langle y^2\rangle^{1/2}= {(2x/p)^{1/p}\over l^{(1-p)/p}} \left({v_l\over V_A}\right)^{m/p},
\label{eq:diffuse2a}
\ee
which implies an outflow layer width of
\be
\langle y^2\rangle^{1/2}\sim l\left({L_x\over l}\right)^{1/p}
\left({v_l\over V_A}\right)^{m/p},
\label{eq:diff2a}
\ee
when $l>L_x$ and 
\be
\langle y^2\rangle^{1/2}\sim \left(L_x l\right)^{1/2}
\left({v_l\over V_A}\right)^{m/p},
\label{eq:diff3a}
\ee
when $L_x>l$ (compare to (\ref{eq:diff2}) and (\ref{eq:diff3})).
Therefore the upper limit on $V_{rec}$ imposed by large scale
field line diffusion is given not by (\ref{eq:lim2a}) but
\be
V_{rec}<V_A\min\left[\left({L_x\over l}\right)^{1/p-1},
\left({l\over L_x}\right)^{1/2}\right]
\left({v_l\over V_A}\right)^{m/p},
\label{eq:lim2aa}
\ee
which shows that for a generous range of $p$ and $m$ the outflow of
fluid from the reconnection zone does significantly limit reconnection
speeds.

In order to recover limits on $p$ or $m$ which are consistent with
fast reconnection we need to consider effects which might provide
more stringent constraints on the reconnection speed.  Some of our
discussion cannot be reproduced for the general case without a
model for the transfer of energy, as well as a scaling of 
eddy anisotropy.  However, one important limit comes from considering
the limits on small scale reconnection events.  
For the general case we replace equation (\ref{eq:vrec1}) with
\be
V_{rec,local}\sim V_A\left({\eta\over V_Al}\right)^{1-p\over2-p} 
\left({v_l\over V_A}\right)^{m/(2-p)}.
\label{eq:vrec1a}
\ee
Following the line of reasoning used in the main text requires 
that the probability a given magnetic field line will return back
to the reconnection zone is small. Formally this means that $p<1$. However,
for $p=1$ the local reconnection speed has already reached 
its maximal value $V_A$, since in the case of locally isotropic eddies the
reconnection geometry is basically the one proposed by Petschek (1964).  
For larger values of $p$ reconnection proceeds at $V_A$. Since $p>1$
corresponds to a spectrum in which eddies become increasingly 
elongated {\it across} magnetic field lines on small scales, this
choice is almost certainly unphysical as well. 

For $p<1$ the Ohmic diffusivity provides the following estimate for
the reconnection rate (compare with equation (\ref{gglob})
\be
V_{rec,global}=k_{\|}L_x V_{rec,local}=V_A
\left({V_Al\over\eta}\right)^{(2p-1)/(2-p)}
{L_x\over l} \left({v_l\over V_A}\right)^{3m/(2-p)},
\label{ggloba}
\ee 
which shows that the reconnection velocity does not depend on
Lundquist number provided that $p>1/2$. Otherwise Ohmic diffusion
will
limit the reconnection rate to the value given by eq.~(\ref{ggloba}).
This will reach the Sweet-Parker reconnection rate when $p=0$,
that is, when the parallel wavelength does not increase as we
go down the turbulent cascade.  Physically, this is equivalent
to assuming that the magnetic field lines are infinitely stiff,
even when the local kinetic energy density is comparable to
the magnetic field energy density.

We note that even taking the relatively modest value of $p=1/2$
implies that the magnetic field suppresses bending motions
far more efficiently than one would suppose from simple
energetic arguments. 

Finally, it is useful to see how these arguments would change
if the Kraichnan spectrum turns out to be correct.  We have
earlier explained why we do not expect this to be the case,
but this is the default model for many researchers.  Also, 
it turns out that this model cannot be found as a special case 
of equation (\ref{k_p2a}).  Instead, while most of the power
is contained in modes with $k_{\perp}\sim k_{\|}$, the amplitude
of the waves is given by
\be
{v_k\over\omega_A}\approx k^{-1}{v_k\over V_A}\ll k^{-1},
\label{ik1}
\ee
which sets this case apart from the strongly nonlinear cascade,
where the wave amplitude is $\sim k_{\perp}^{-1}$.  If
we combine the assumption of isotropy in wavevector space
with equations (\ref{tau}) and (\ref{eq:casc}) we find
the usual result, for the Kraichnan spectrum, that
\be
v_k\sim v_l (kl)^{-1/4}.
\label{ik2}
\ee
Combining this with equation (\ref{ik1}) we see that the characteristic
displacement on a scale $k^{-1}$ is
\be
<y^2>^{1/2}\sim {v_l\over V_A} {l\over (kl)^{5/4}}.
\ee
This implies a field line diffusion, analogous to 
equation (\ref{eq:diffuse}) of
\be
{d\langle y^2\rangle\over dx}\sim l
\left({\langle y^2\rangle\over l^2}\right)^{3/4}
\left({v_l\over V_A}\right)^2,
\ee
which in turn leads to an outflow layer width, for $L_x<l$, of
\be
\langle y^2\rangle^{1/2}\sim l\left({L_x\over l}\right)^2
\left({v_l\over V_A}\right)^4.
\ee
The upper limit on $V_{rec,global}$ set by field line diffusion
is therefore
\be
V_{rec,global}<V_A {L_z\over l}\left({v_l\over V_A}\right)^4.
\label{ik3}
\ee

To see that this is the actual speed of reconnection we
need to evaluate $V_{rec,local}$.  The Kraichnan spectrum
will be truncated due to resistivity at a wavenumber given
by
\be
k_c \sim l^{-1} \left({v_l\over V_A}\right)^{4/3} 
\left({V_Al\over \eta}\right)^{2/3}.
\ee
Using the usual Sweet-Parker formula for the local reconnection
speed we get
\be
V_{rec,local}\sim V_A \left({v_l\over V_A}\right)^{2/3} 
\left({\eta\over V_Al}\right)^{1/6}.
\ee
We can obtain the same estimate by ignoring the spectral cutoff,
and calculating the appropriate value of $k$ by requiring that
$\eta \langle y^2\rangle^{-1/2}$ equal the local Sweet-Parker
reconnection speed.  The upper limit on the global reconnection
speed is recovered by multiplying $V_{rec,local}$ by $kL_x$.
We obtain
\be
V_{rec,global}<V_A
\left({V_Al\over\eta}\right)^{1/2}
{L_x\over l} \left({v_l\over V_A}\right)^4.
\label{ik4}
\ee
Comparing equations (\ref{ik3}) and (\ref{ik4}) we see that in
most cases the reconnection speed will be given by equation
(\ref{ik3}).  In other words, as usual, global field line
diffusion provides the real constraint on magnetic reconnection in
this case.

\pagebreak

\begin{figure}
\begin{picture}(441,216)
\includegraphics{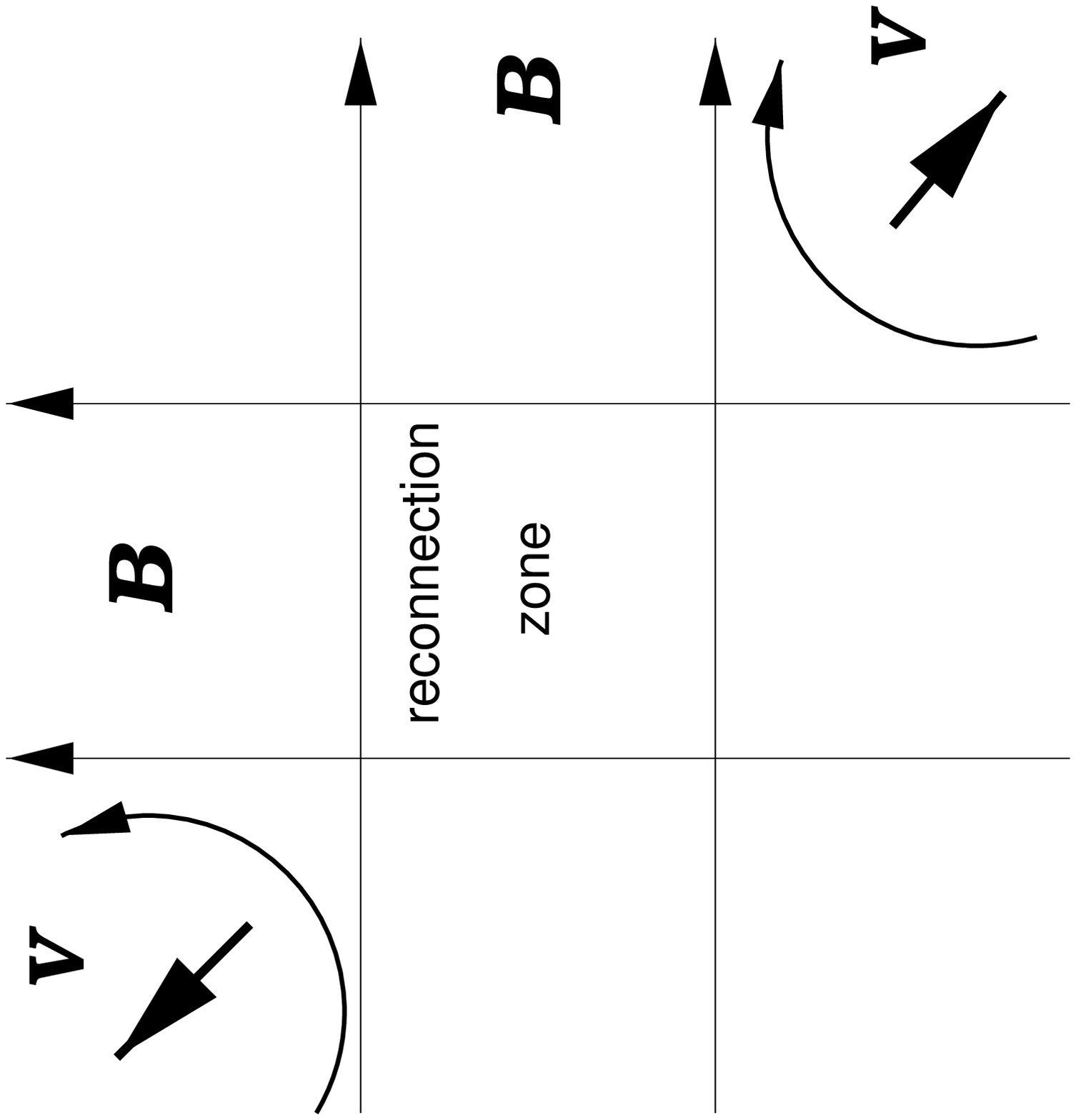}
\end{picture}
\caption[]{The geometry of magnetic field lines in 3D reconnection.
The reconnected lines stretch and carry the conducting plasma with them.
The plasma is also redistributed along the field lines.}
\end{figure}

\clearpage

\begin{figure}
\begin{picture}(441,216)
\includegraphics{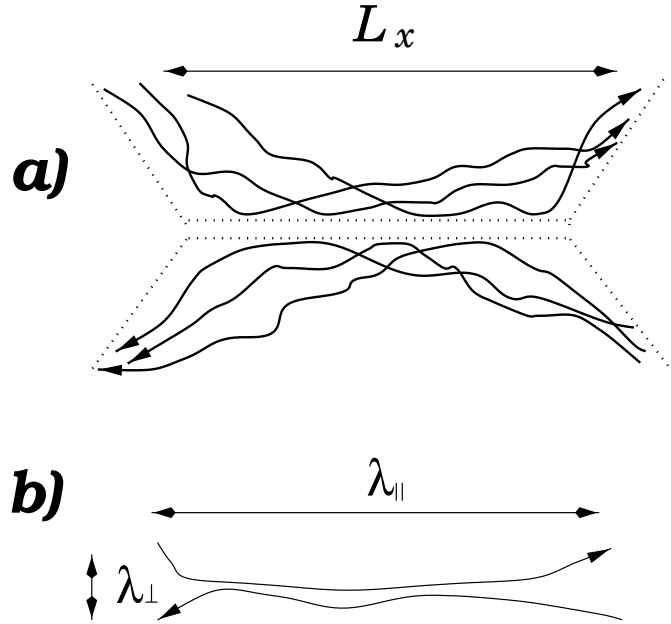}
\end{picture}
\caption[]{ a) The structure of the reconnection region when the field is
turbulent. Local reconnection events happen on the small scale $\lambda_{\|}$
rather than $L_x$ and this accelerates reconnection. The plasma
is redistributed along the field lines in a layer of thickness
$\langle y^2\rangle^{1/2}$, which is much thicker than the region $\sim \lambda_{\bot}$
from which the ejection of the magnetic field takes place. b) Local structure
of magnetic field lines.}
\end{figure}

\end{document}